\documentclass[aps,prl,twocolumn,letterpaper,superscriptaddress,showpacs]{revtex4-1}
\usepackage{graphicx}
\usepackage{CJK}
\usepackage{verbatim}
\usepackage{physics}
\usepackage[colorlinks,linkcolor=blue,anchorcolor=blue, citecolor=blue,urlcolor=blue,]{hyperref}
\usepackage{lineno} 
\usepackage{bm}
\usepackage{mathptmx}
\usepackage{multirow}
\makeatletter

\begin{document}
\begin{CJK*}{UTF8}{bsmi}

\title{Pressure-induced trans-proximate correlation in La$_4$Ni$_3$O$_{10}$  \\and possible routes to enhance its superconductivity}

\author{Ruoshi Jiang}
\affiliation{Department of Materials Science and Metallurgy, University of Cambridge, Cambridge CB3 0FS, United Kingdom}
\affiliation{School of Physics and Astronomy \& Tsung-Dao Lee Institute, Shanghai Jiao Tong University, Shanghai 200240, China}
\author{Zhiyu Fan }
\affiliation{School of Physics and Astronomy \& Tsung-Dao Lee Institute, Shanghai Jiao Tong University, Shanghai 200240, China}
\author{Bartomeu Monserrat}
\affiliation{Department of Materials Science and Metallurgy, University of Cambridge, Cambridge CB3 0FS, United Kingdom}
\author{Wei Ku}
\altaffiliation{corresponding email: weiku@sjtu.edu.cn}
\affiliation{School of Physics and Astronomy \& Tsung-Dao Lee Institute, Shanghai Jiao Tong University, Shanghai 200240, China}
\affiliation{Key Laboratory of Artificial Structures and Quantum Control (Ministry of Education), Shanghai 200240, China}
\affiliation{Shanghai Branch, Hefei National Laboratory, Shanghai 201315, People's Republic of China}

\begin{abstract}
We report an unexpected trans-proximate interlayer correlation (stronger correlation between disjoint layers than the adjacent ones) in the high-pressure phase of the recently discovered La$_4$Ni$_3$O$_{10}$ superconductors.
Accompanied by an unusual pressure-induced fractionalization of Ni$^{2+}$ ionic spin from the standard spin-1 to spin-$\frac{1}{2}$, this trans-proximate correlation results from the emergence of a cross-layer trimer in our multi-energy-scale derivation of the electron dynamics.
The resulting low-energy effective description resembles that of the cuprates and suggests a universal superconducting mechanism in all existing nickelate and cuprate superconductors.
The rare trans-proximate correlation not only explains the weaker superconductivity in comparison with the related La$_3$Ni$_2$O$_7$ samples, but it also indicates a viable strategy to improve superconductivity in this trilayer nickelate by lowering layer symmetry.
Such pressure-induced trans-proximate correlation is expected in many materials and examplifies the engineering of rich uncharted quantum states of matter through pressure.
\end{abstract}
\maketitle
\end{CJK*}

Applied pressure is known to be a clean and yet effective mean to change the physical properties of materials~\cite{Dagotto2005} or even to alter quantum states of matter, such as Mott insulator\,\cite{Yang2023Pressure-induced}, charge density wave\,\cite{Tzong-Ru2015Exploration}, or magnetic ordering\,\cite{Sheng1996Pressure-induced}. 
A notable example is the pressure-modulation of unconventional superconductivity in the cuprates\,\cite{Zhou2022Quantum}, iron-pnictides\,\cite{Zapf2016Europium, Sun2015Dome-shaped, Reiss2020Quenched}, nickelates\,\cite{Wang2022Pressure-induced}, manganites\,\cite{Dissanayake2023Helical},  ruthenates\,\cite{Alireza2010Evidence}, vanadates\,\cite{Yu2021Unusual, Du2021Pressure}, titenates\,\cite{Nie2023Pressure}, and UTe$_2$\,\cite{Ran2021Expansion}.
In these materials, the superconductivity is often \textit{quantitatively} strengthened under moderate pressure through enhanced kinetics of itinerant carriers.

Recently, a more exotic \textit{qualitative} effect of pressure is observed in the bilayer nickelate superconductor\,\cite{Hualei0516}, La$_3$Ni$_2$O$_7$, which abruptly develops superconductivity above $14$\,GPa pressure, with a remarkable $80$\,K transition temperature\,\cite{Hualei0516,Hou0719,Yanan0727}.
Unlike the first class of nickelate superconductors that is chemically engineered to contain spin-$\frac{1}{2}$ Ni$^{+}$ ions\,\cite{Berit2021Doping, Ding2024Cuprate-like, lang2021, lang2022} similar to Cu$^{2+}$ in the cuprates\,\cite{Azuma1992Superconductivity}, this new class has an unexpected nominal spin-1 Ni$^{2+}$ ionic configuration.
As a result, it only develops the observed high-temperature superconductivity in the high pressure phase, in which the ionic spin is unusually fractionalized from 1 to $\frac{1}{2}$ below the eV scale\,\cite{Jiang2024}.
Such an alteration of correlation opens up a new class of pressure effects ideal for exploring uncharted quantum states of matter.


In this context, the most recent discovery of superconductivity around $30$\,K in trilayer La$_4$Ni$_3$O$_{10}$ beyond a pressure of $14$-$20$\,GPa\,\cite{Sakakibara2024Theoretical, Zhu2024Superconductivity, Li2024Signature, Zhang2024Superconductivity, Li2024Structural} offers a valuable comparison for the physical properties and the corresponding electronic structures relevant to unconventional high-temperature superconductivity\,\cite{Carvalho2000Influence,Wu2001Magnetic,Li2017Fermiology,Zhang2020High,Susmit2020Physical,Li2020Metal-to-metal,Huang2020Correlation,Rout2020Structural,Zhang2020Interwined,Sakakibara2024Theoretical,Ning2024High-pressure, Zhu2024Superconductivity, Zhang2024Superconductivity, Li2024Signature, Li2024Structural, Kakoi2024Multiband, Leonov2024Electronic, Wang2024Non-Fermi, Tian2024Effective, LaBollita2024Electronic, Yang2024Effective, Chen2024Trilayer, Du2024correlated, Xu2024Origin, Qin2024Frustrated,Li2025Distinct, Zhang2024Prediction, Sonia2024Dynamics, Zhang2025Bulk, Mingzhe2025Direct, Hu2025Origin, Di2025Isotropic, khasanov2025identical, Xu2025Collapse}.
For example, both trilayer and bilayer nickelates exhibit pressure-induced structural and superconducting transitions, and at high pressure their transport properties switch from Fermi liquid to non-Fermi liquid\,\cite{Zhu2024Superconductivity, Zhang2024Superconductivity}.
Therefore, it is of timely and significant interest to investigate the pressure effects on the low-energy electronic structure, particularly focusing on the possible alteration of electronic correlation ideal for unconventional superconductivity, in comparison with the bilayer La$_3$Ni$_2$O$_7$.

Here, we report yet another exotic pressure-induced qualitative change of electronic correlation in La$_4$Ni$_3$O$_{10}$, namely a trans-proximate correlation across layers (i.e. stronger correlation between disjoint layers than the adjacent ones).
Accompanying the emergence of a cross-layer trimer in the high-pressure phase at sub-eV scale from our multi-energy-scale analysis, this rare correlation naturally explains the weaker superconductivity in this material by limiting the fluctuation of the fractionalized Ni$^{2+}$ ionic $\frac{1}{2}$-spin.
Our finding suggests a route for enhancing superconductivity in trilayer nickelates through lowering their layer symmetries and thus unifying the superconducting mechanism with previous nickelate and cuprate superconductors.
In general, the discovered exotic trans-proximate correlation can be expected in many functional materials under applied pressure to produce rich classes of uncharted quantum states of matter.

We start by considering the Hartree scale physics of charge distribution to identify the location of the self-doped carriers and the associated valence of the Ni ions.
To explore this high energy regime in which inter-atomic magnetic correlations are negligible, we evaluate the one-body spectral function arising from carrier correlation with \textit{unordered} Ni ionic spins in the Curie paramagnetic phase\,\cite{Jiang2022,Jiang2024}, with full computational details in the SI\,\cite{supplementary}.
In short, we use the all-electron implementation of density functional theory\,\cite{DFT1, DFT2} in {\sc Wien2k}\,\cite{Blaha2001, Blaha2019} within the LDA+$U$ formalism\,\cite{Anisimov1993,Liechtenstein1995} with a typical $U=6$\,eV for nickel $d$-orbitals\,\cite{Hualei0516, Dudarev1998}. 
We note that our conclusions are insensitive to the value of $U$ within a few eV\,\cite{Jiang2024Interactionan}.
For computational efficiency, we subsequently employ an effective interacting Hamiltonian $H^{(\mathrm{Hartree})}$\,\cite{supplementary,lang2021,Jiang2022, Jiang2024} obtained from symmetry-preserving Wannier functions\,\cite{Wei2002, Marzari1997}.

\begin{figure}
\centering
\includegraphics[width=0.85\columnwidth]{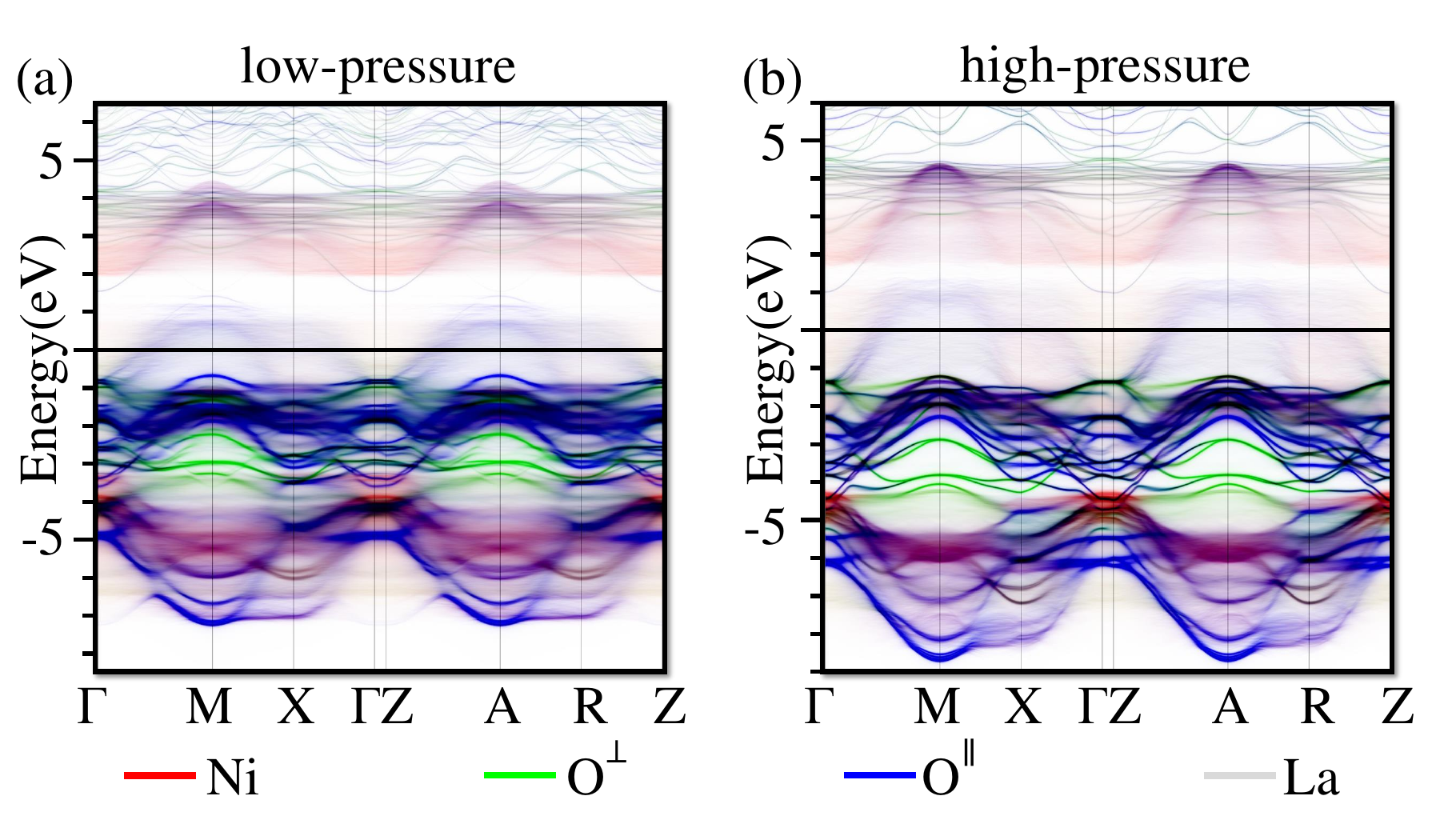}
\vspace{-0.4cm}
\caption{
\textit{Itinerant carriers and their correlation with ionic moments.}
One-body spectral functions of Curie-paramagnetic La$_4$Ni$_3$O$_{10}$ in the (a) low-pressure and (b) high-pressure phases. In both phases, charge carriers are holes on the in-plane oxygens (O$^\parallel$) that strongly scatter with adjacent Ni ionic spins, leading to the observed spread.
}
\vspace{-0.6cm}
\label{fig1}
\end{figure}

Figure~\ref{fig1} shows the one-body spectral functions of the low-pressure and high-pressure phases, unfolded\,\cite{Wei2010} to the smaller high-pressure unit cell containing three nickel atoms, and using the experimental crystal structures at ambient pressure and 44.3\,GPa\,\cite{Li2024Structural}.
Only the bands associated with O$^\parallel$-$p$ orbitals in the Ni layer (blue) cross the Fermi energy, indicating the presence of a single dominant type of carrier in the system, similar to the bilayer nickelates.
We also note that the negligible contribution of the out-of-plane apical O$^\perp$ orbitals (green) near the Fermi energy indicates that the charge fluctuations of itinerant carriers to O$^\perp$ are negligible.
This result unambiguously establishes that the effective charge carriers reside mostly in the in-plane O orbitals, as in the cuprate\,\cite{Zhang1988}, NdNiO$_2$\,\cite{lang2021,lang2022}, and La$_3$Ni$_2$O$_7$\,\cite{Jiang2024} superconductors.
Furthermore, the one-body spectral function indicates that the occupation of the in-plane oxygen O$^\parallel$ bands crossing the Fermi level is approximately uniform across all three layers in both low-pressure and high-pressure phases.

The Ni bands (red) in Fig.\,\ref{fig1} result from the ensemble average of configurations each having 8 occupied and 2 unoccupied Ni-$d$ bands, corresponding to a spin-polarized Ni$^{2+}$-$d^8$ ionic configuration.
Their broad spectral distribution reflects the strong scattering associated with spatial fluctuations of their spin orientation.
Furthermore, the Ni-$d$ orbitals make a negligible contribution near the Fermi energy, indicating their weak charge fluctuations.
Therefore, only the spin and corresponding fluctuation of the Ni-$d$ bands are active at low energy and correlate with the itinerant carriers on the O$^\parallel$-$p$ bands.

Importantly, correlation with Ni spins strongly influences the motion of carriers, as clearly manifested in their broad quasi-particle peaks in Fig.\,\ref{fig1}. 
As a comparison, observe the sharp quasi-particle peaks for orbitals that correlate weakly with the Ni ions, such as La (in gray) or the outermost apical O of the trilayer (in green).
Consistently, the carrier bandwidth of about $2$\,eV in Fig.\,\ref{fig1} is also heavily renormalized from the bare bandwidth of about $3$\,eV in a non-magnetic or magnetically coherent phase\,\cite{supplementary, Wang2024Non-Fermi}, consistent with experimental observations\,\cite{Li2017Fermiology,Xu2024Origin}.
The significant mass enhancement and spectral broadening of quasi-particles reflect the intense scattering of the carriers due to their strong correlation with the unordered magnetic moments of Ni ions.

\begin{table}
    \vspace{-0.1cm}
    \caption{
    \textit{Pressure changes at Hartree scale: dramatic enhancement of $t^\perp_{pZ}$.}
    DFT-derived in-plane ($\parallel$) and out-of-plane ($\perp$) leading hopping parameters in $H^{(\text{Hartree})}$ within and between the middle ($\text{m}$) and lower ($\text{l}$) layers, in unit of eV.
    Subscripts $p$, $X$, and $Z$ denote the O-$p$, Ni-$d_{x^2-y^2}$, and Ni-$d_{3z^2-r^2}$ orbitals, respectively.
    }
\begin{ruledtabular}
\begin{tabular}{c@{\hspace{0.5cm}}|ccccc}
$H^{(\text{Hartree})}$ & $t^{\parallel (\text{mm})}_{pX}$ & $t^{\parallel(\text{ll})}_{pX}$ &  $t^{\perp (\text{mm})}_{pZ}$ & $t^{\perp (\text{ml})}_{pZ}$   & $t^{\perp (\text{ll})}_{pZ}$\\ \hline
$\mathrm{low\mbox{-}pressure}$  & 1.43       & 1.42       & \textbf{1.38}       & 1.76       & 1.21 \\ 
$\mathrm{high\mbox{-}pressure}$ & 1.76       & 1.76       & \textbf{1.94}       & 2.00       & 1.62 \\
\end{tabular}
\end{ruledtabular}
\label{tab1}
\vspace{-0.5cm}
\end{table}

Table~\ref{tab1} gives the leading hopping parameters obtained from our Wannier function analysis\,\cite{supplementary} in both low- and high-pressure phases.
Moving from low to high pressure dramatically enhances the inter-plane hopping $t^\perp_{pZ}$, between the Ni-$d_{3z^2-r^2}$ and the O$^\perp$-$p_z$ orbitals sandwiching the middle layers, by $40$\% from $1.38$\,eV to $1.94$\,eV, as highligted in bold in Tab.\,\ref{tab1}.
As a comparison, the most relevant hopping parameter of the itinerant carriers, $t^\parallel_{pX}$ between Ni-$d_{x^2-y^2}$ and in-plane O$^\parallel$-$p_x$/$p_y$ orbitals, only increases by about $20$\% in moving from low to high pressure.

The large enhancement in $t^\perp_{pZ}$ is likely associated with qualitative changes of the low-energy electronic structure at high pressure.
A possible scenario could be the formation of strong bonding of Ni-$d_{3z^2-r^2}$ orbitals across layers\,\cite{Sakakibara2024Theoretical, LaBollita2024Electronic, Chen2024Trilayer, Du2024correlated, Zhang2024Prediction} via a large $t^\perp_{pZ}$ that overcomes the charge transfer energy $E_\mathrm{CT}=E(\ket{d^9 p^5 d^8})-E(\ket{d^8 p^6 d^8})$.
However, Fig.\,\ref{fig1} shows that $E_\mathrm{CT}$, the energy difference between the unoccupied Ni-$d_{3z^2-r^2}$ orbital (red) and the occupied O$^\perp$-$p_z$ orbital (green), is in the range $3$-$4$\,eV.
This indicates that even in the high-pressure phase this system is far from the $t^\perp_{pZ} \gg E_\mathrm{CT}$ bonding limit, making strong bonding of Ni-$d_{3z^2-r^2}$ orbitals across layers implausible.
To identify the driving mechanism behind the pressure-driven changes, we next explore lower-energy scales.

First, we note that the enhancement of the trilayer Ni-O-Ni-O-Ni coupling does not directly contribute to the motion of carriers.
The former concerns the Ni-$d_{3z^2-r^2}$ and O$^\perp$-$p_z$ orbitals across layers, while the latter mostly involves the O$^\parallel$-$p_x$/$p_y$ and Ni-$d_{x^2-y^2}$ orbitals in the same layer.
The kinetic processes between these two sub-spaces are heavily suppressed due to the weaker hopping between them and due to the $x^2-y^2$ symmetry of carriers around each Ni (arising from strong coupling to Ni-$d_{x^2-y^2}$ orbitals).
The lack of O$^{\perp}$ (green) contributions to the itinerant bands in Fig.\,\ref{fig1} is a clear indication of such decoupled kinetics.
Therefore, the primary effects of the $t^\perp_{pZ}$ enhancement on the low-energy physics should be through a qualitative change in the local electronic structure of the trilayer Ni-O-Ni-O-Ni component and its correlation with the itinerant carriers.

We thus investigate the \textit{local} many-body electronic structure of the trilayer Ni-O-Ni-O-Ni component, seeking to identify any bifurcating low-energy dynamics that qualitatively distinguishes the low- and high-pressure phases\,\cite{note_hole_impact}.
We start with $H^{(\mathrm{Hartree})}$ in the atomic Wannier basis\,\cite{Wei2002,supplementary},
containing typical intra-atomic Coulomb interactions $U_{d}=6$, $U_{p}=4$\,\cite{Ogata_2008}, $J_{\mathrm{H}}=0.8$\,eV, and $E_\mathrm{CT}\sim 3$\,eV. We derive the associated eV-scale effective Hamiltonian within the local $\ket{d^8p^6d^8p^6d^8}$ subspace by numerically integrating out states containing double occupied orbitals of Ni. 

Specifically, this is accomplished by numerically applying a series of \textit{symmetric} unitary transformations to completely zero out the coupling between the high-energy states from the remaining low-energy subspace\,\cite{supplementary}, effectively absorbing the short-time quantum fluctuations into the dressing of the longer-lived low-energy objects.
Other than taking special care on explicitly maintaining the symmetry of the representation, this numerical procedure is a straightforward implementation of the numerical canonical transformation\,\cite{White2002}, which itself can be considered a complete \textit{non-perturbative} extension of the Schrieffer-Wolff transformation\,\cite{Schrieffer1966,Zaanen1988,lang2021,Yin2009}.
We confirm the completeness of the resulting effective description by comparing its eigen-solutions against the low-energy ones of the original Hamiltonian.

The fully occupied O$^\perp$-$p$ shell obtained upon integrating out states with double occupation of Ni orbitals implies that among the local $\ket{d^8p^6d^8p^6d^8}$ states the remaining low-energy subspace has no charge freedom left.
The resulting eV-scale dynamics is therefore that of six coupled \textit{dressed} spins residing in the Ni $e_g$ orbitals:
\begin{align}
H^{\mathrm{(eV)}} =&-\Tilde{J}^{(\text{m})}_\mathrm{H}~\mathbf{S}_X^{(\text{m})}\cdot\mathbf{S}_Z^{(\text{m})}
-\Tilde{J}^{(\text{l})}_\mathrm{H}~(\mathbf{S}_X^{(\text{u})}\cdot\mathbf{S}_Z^{(\text{u})}
+\mathbf{S}_X^{(\text{l})}\cdot\mathbf{S}_Z^{(\text{l})}) \nonumber \\
&+ J_{ZZ}~(\mathbf{S}_Z^{(\text{m})}\cdot\mathbf{S}_Z^{(\text{u})}
+ \mathbf{S}_Z^{(\text{m})}\cdot\mathbf{S}_Z^{(\text{l})}), 
\label{eq_6spin}
\end{align}
where in all Ni$^{2+}$ ions belonging to the upper (u), middle (m), and lower (l) layers, the ferromagnetic intra-atomic Hund's coupling between spins in the $d_{x^2-y^2}$ orbital, $\mathbf{S}_X$, and $d_{3z^2-r^2}$ orbital, $\mathbf{S}_Z$, is slightly renormalized by the higher-energy charge fluctuation to $\Tilde{J}_\mathrm{H}\sim 0.6$\,eV from its bare value $J_\mathrm{H}\sim 0.8$\,eV .
In addition, an interlayer anti-ferromagnetic super-exchange\,\cite{Anderson1950}, $J_{ZZ}$, emerges between Ni ions.

\begin{table}
    \vspace{-0.3cm}
    \caption{
    \textit{Pressure changes at eV scale: superexchange coupling $J_{ZZ}$.}
    Emergent interlayer superexchange $J_{ZZ}$ and renormalized Hund's couplings $\tilde{J}_{\mathrm{H}}$ in the middle ($\text{m}$) and lower ($\text{l}$) layers in $H^\mathrm{(eV)}$, in units of eV.
}
    \begin{ruledtabular}
        \begin{tabular}{c @{\hspace{0.5cm}}|cccc}
$H^{\mathrm{(eV)}}$&$\tilde{J}^{(\text{m})}_{\mathrm{H}}$ & $\tilde{J}^{(\text{l})}_{\mathrm{H}}$&& $J_{ZZ}$  \\ 
        \hline
$\mathrm{low\mbox{-}pressure}$   & \textbf{0.63} & \textbf{0.65}& $>$ & 0.36   \\ 
$\mathrm{high\mbox{-}pressure}$  & 0.57 & 0.62& $\mathbf{<}$ & \textbf{0.66}   \\
        \end{tabular}
    \end{ruledtabular}
    \label{tab2}
    \vspace{-0.5cm}
\end{table}

Interestingly, Tab.\,\ref{tab2} shows an important distinction in $H^{(\mathrm{eV})}$ between the low- and high-pressure phases of trilayer La$_4$Ni$_3$O$_{10}$, as previously also observed in the bilayer nickelates~\cite{Jiang2024}.
At low pressure, even though a rather strong anti-ferromagnetic inter-atomic super-exchange coupling, $J_{ZZ}\sim0.4$\,eV, emerges between the $d_{3z^2-r^2}$ orbitals, it is, as expected, weaker than the ferromagnetic intra-atomic Hund's coupling $\Tilde{J}_\mathrm{H}\sim 0.6$\,eV.
By contrast, at high pressure, the enhanced $t^\perp_{pZ}$ leads to a highly unusual situation in which the \textit{interlayer} super-exchange $J_{ZZ}\sim0.7$\,eV is stronger than the intra-atomic Hund's coupling $\tilde{J}_\text{H}\sim0.6$\,eV.
Note that inclusion of additional fluctuations of itinerant carriers to the $d_{x^2-y^2}$ orbital can only further weaken $\tilde{J}_\mathrm{H}$ and in turn enhance the high-pressure behavior\,\cite{supplementary}.
This switching of dominant couplings between the low- and high-pressure regimes is the direct indication we seek for a qualitative change of the low-energy physics. 

To explicitly demonstrate the qualitatively distinct physical behavior, we drive the local physics in both phases to an even lower sub-eV scale by repeating the above numerical canonical transformation\,\cite{supplementary}.
At low pressure (LP), the canonical transformation of $H^{\mathrm{(eV)}}$ that brings the dominant Hund's coupling $\tilde{J}_\mathrm{H}$ terms to a diagonal form, gives an $3+1$ eigen-structure with the intra-atomic spin-0 singlet as the higher-energy object to be integrated out.
Upon a further numerical canonical transformation to fully decouple this singlet from the lower-energy subspace, only an effective spin-1 triplet $\mathbf{S}_{\mathrm{eff}}$ in each Ni ion remains in the slower sub-eV-scale local dynamics:
\begin{equation}
H_{\text{LP}}^{\mathrm{(sub\text{-}eV)}} = J^{(\text{ml})}~(\mathbf{S}_{\mathrm{eff}}^{(\text{m})}\cdot\mathbf{S}_{\mathrm{eff}}^{(\text{u})}+\mathbf{S}_{\mathrm{eff}}^{(\text{m})}\cdot\mathbf{S}_{\mathrm{eff}}^{(\text{l})}),
    \label{SubeV_LP}
\end{equation}
and they couple to each other through a renormalized antiferromagnetic super-exchange $J^{(\text{ml})}$. 
These spin-1 ions coupled with large $J^{(\text{ml})}\sim 0.11$\,eV (c.f. Tab.~\ref{tab3}) result in strong interlayer anti-ferromagnetic correlations and strongly promote magnetic ordering (unless overwhelmed by itinerant carrier-induced long-range fluctuations\,\cite{Tan2022}).
Indeed, current experimental studies, including magnetic susceptibility and resistivity\,\cite{Zhu2024Superconductivity,Li2024Signature}, X-ray diffraction\,\cite{Zhang2020High, Susmit2020Physical,Li2020Metal-to-metal, Huang2020Correlation, Li2024Structural, Zhang2020Interwined}, neutron diffraction\,\cite{Zhang2020Interwined, Zhu2024Superconductivity}, Raman\,\cite{Sonia2024Dynamics}, nuclear magnetic resonance\,\cite{Kakoi2024Multiband}, and ultrafast spectroscopy\,\cite{Xu2025Collapse}, all have identified signals consistent with anti-ferromagnetic order.

\begin{table}
    \vspace{-0.3cm}
    \caption{
    \textit{Pressure induced fractionalization of ionic spins and emergence of a cross-layer trimer composite spin.}
    Derived effective Ni$^{2+}$ ionic spins and their coupling in $H^{\mathrm{(sub\text{-}eV)}}$ are given (in unit of eV).
    The introduction of an O vacancy greatly suppresses the interlayer couplings at high pressure.
    }
    \begin{ruledtabular}
\begin{tabular}{c@{\hspace{0.3cm}}|cccc@{\hspace{0.3cm}}|c}
\multicolumn{1}{c|}{$H^{\mathrm{(sub\text{-}eV)}}$ }&\multicolumn{4}{c|}{} &\multicolumn{1}{c}{with O vacancy}\\ \hline
$\mathrm{low\mbox{-}pressure}$ &$J^{(\text{ml})} $ &&&& $J^{(\text{ml})}$   \\ \cline{2-6}
$S_{\mathrm{eff}}=1$ &  0.11 &&&& 0.13  \\ \hline
$\mathrm{high\mbox{-}pressure}$ & $J^{(\text{lt})}$ &$J^{(\text{mt})}$ & $K^{(\text{lt})}$ & $K^{(\text{mt})}$& $J^{(\text{ml})}$   \\ \cline{2-6}
$S_{\mathrm{eff}}=1/2$  & $\mathbf{-0.28}$ & 0.13 & \textbf{0.15} &$-$0.09& \textbf{0.08} \\
\end{tabular}
\end{ruledtabular}
\label{tab3}
\vspace{-0.4cm}
\end{table}

\begin{figure}
\centering
\vspace{-0.6cm}
\includegraphics[width=0.95\columnwidth]{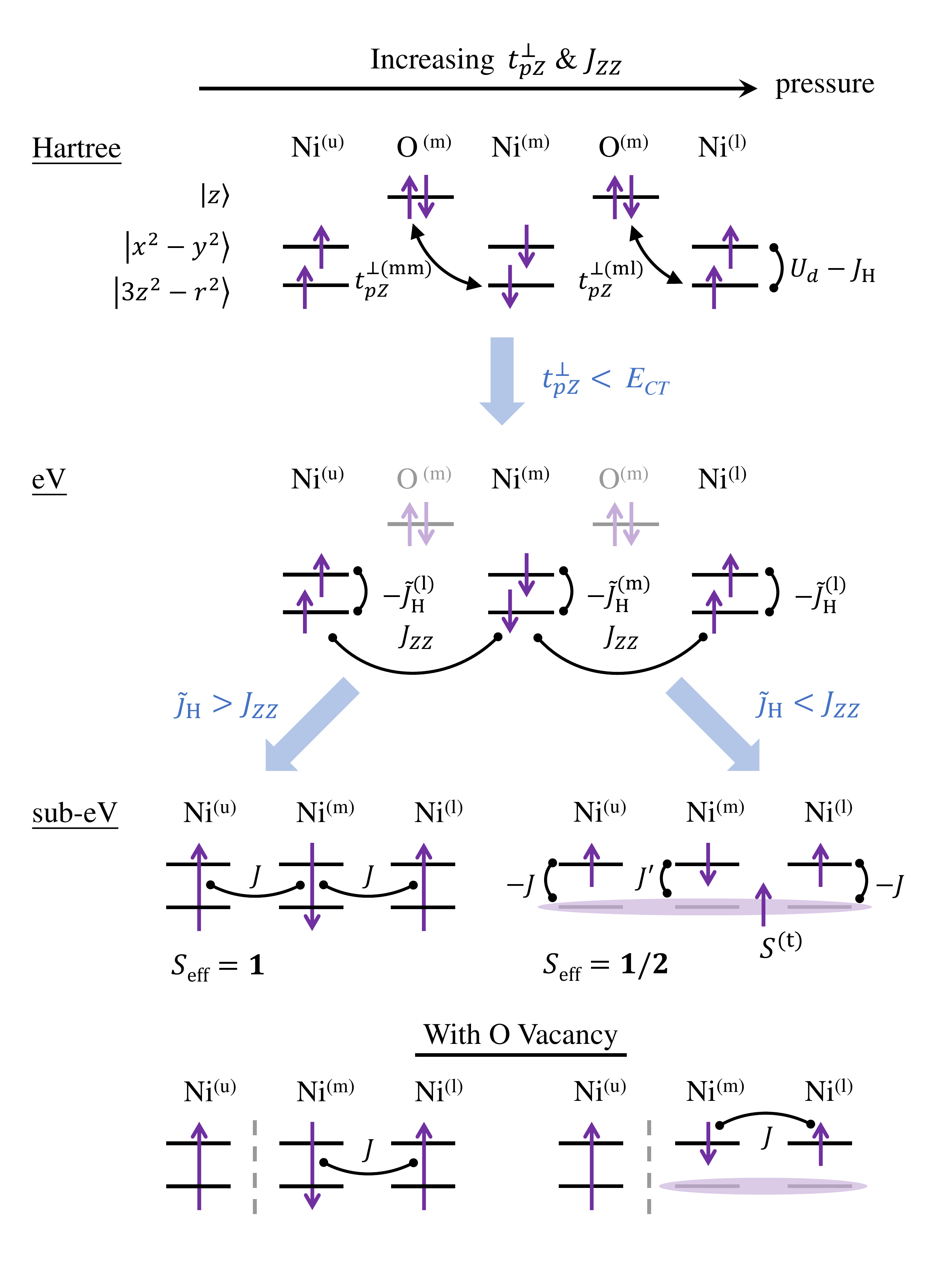}
\vspace{-0.5cm}
\caption{\textit{Multi-scale illustration of the pressure-induced fractionalization of ionic spins and emergence of cross-layer trimer composite spins in La$_4$Ni$_3$O$_{10}$.}
Starting from Hartree scale, upon absorbing rapid charge fluctuations of Ni$^{2+}$ ions in the local Ni-O-Ni-O-Ni trilayer component, the fully occupied \textit{effective} O$^{(\text{m})}$-$p_z$ orbital becomes inert (shadowed), leaving only spins dynamics of Ni orbitals active at eV scale.
At low-pressure, the stronger effective Hund's coupling $\Tilde{J}_\mathrm{H}$ dictates a typical spin-1 Ni$^{2+}$ ions at sub-eV scale with large interlayer exchange energy.
In contrast, due to the unusually strong interlayer super-exchange $J_{ZZ}$ at high-pressure, a cross-layer trimer composite $\frac{1}{2}$-spin (purple ellipsoid) emerges from spins in the $d_{3z^2-r^2}$ orbitals and effectively fractionalizes the Ni$^{2+}$ ionic spin from 1 to $\frac{1}{2}$ at sub-eV scale.
To enhance the supercondcuting phase stiffness through relieving the trimer-induced interlayer correlation, a possible scenario is to promote singlet dimerization via lowering trilayer symmetry, for example with a O vacancy.
}
\vspace{-0.6cm}
\label{fig2}
\end{figure}

By contrast, for the high-pressure (HP) phase the dominant interlayer anti-ferromagnetic super-exchange $J_{ZZ}$ dictates a strong correlation between $\mathbf{S}_Z$ in the trilayer.
Indeed, the numerical canonical transformation of $H^{(\mathrm{eV})}$ to bring the dominant super-exchange $J_{ZZ}$ terms to a diagonal form gives a well separated $2+2+4$ eigen-structure, with the lowest-energy doublet being a cross-layer trimer that emerges as a composite $\frac{1}{2}$-spin, $\mathbf{S}^{\mathrm{(t)}}$, see Fig.\,\ref{fig2} for a schematic illustration.
Consequently, just as in the bilayer nickelate\,\cite{Jiang2024}, the Ni$^{2+}$ ionic spin $\mathbf{S}_{\mathrm{eff}}$ is fractionalized to spin-$\frac{1}{2}$ from the remaining $\mathbf{S}_X$.

We have thus successfully identified the key physical impact of applied pressure that gives rise to the \textit{qualitatively} distinct low-energy behaviors of the local trilayer Ni-O-Ni-O-Ni component in La$_4$Ni$_3$O$_{10}$.
As illustrated in Fig.\,\ref{fig2}, in the low-pressure phase, Ni$^{2+}$ ions are in the usual spin-1 $d^8$-configuration dictated by Hund's coupling.
By contrast, in the high-pressure phase, the dramatically enhanced $t^\perp_{pZ}$ promotes a much stronger eV-scale $J_{ZZ}$, which in turn drives the formation of a cross-layer trimer composite $\frac{1}{2}$-spin.
In the sub-eV scale, this effectively fractionalizes the ionic spin of Ni$^{2+}$ from 1 to $\frac{1}{2}$, without dramatically altering their charge.
Other than the addition of interlayer trimer composite spins to be elaborated below, the low-energy effective physics is therefore controlled by in-plane motion of itinerant carriers strongly correlated with the spin-$\frac{1}{2}$ ionic spins, analogous to other nickelate and cuprate superconductors.
This similarity in the building blocks of the low-energy electronic structure strongly suggests a similar microscopic mechanism and ($d$-wave) symmetry as the cuprates\,\cite{Wu2019Robust, Ding2024Cuprate-like, Lechermann2023Electronic, Jiang2024High-Temperature}.

However, distinct from the nearly decoupled bilayer in La$_3$Ni$_2$O$_7$, the emergence of the additional composite $\frac{1}{2}$-spin provides an efficient connection between the three layers in La$_4$Ni$_3$O$_{10}$.
Indeed, upon a further numerical canonical transformation\,\cite{supplementary} to decouple the higher-energy quadruplets and doublets, the resulting sub-eV dynamics of the system is described by:
\begin{align}
H_{\text{HP}}^{\mathrm{(sub\text{-}eV)}} 
&=J^{(\text{mt})}~\mathbf{S}_{\mathrm{eff}}^{(\text{m})}\cdot\mathbf{S}^{(\text{t})}+J^{(\text{lt})}~(\mathbf{S}_{\mathrm{eff}}^{(\text{l})}\cdot\mathbf{S}^{(\text{t})}+\mathbf{S}_{\mathrm{eff}}^{(\text{u})}\cdot\mathbf{S}^{(\text{t})}) \nonumber\\
&+\hspace{-0.05cm}K^{(\text{lt})}\hspace{-0.05cm}\Big[(\mathbf{S}_{\mathrm{eff}}^{(\text{m})}\cdot\mathbf{S}_{\mathrm{eff}}^{(\text{u})})(\mathbf{S}_{\mathrm{eff}}^{(\text{l})}\cdot\mathbf{S}^{(\text{t})})\hspace{-0.05cm}+\hspace{-0.05cm}(\mathbf{S}_{\mathrm{eff}}^{(\text{m})}\cdot\mathbf{S}_{\mathrm{eff}}^{(\text{l})})(\mathbf{S}_{\mathrm{eff}}^{(\text{u})}\cdot\mathbf{S}^{(\text{t})})\Big] \nonumber\\
&+\hspace{-0.05cm}K^{(\text{mt})}~(\mathbf{S}_{\mathrm{eff}}^{(\text{u})}\cdot\mathbf{S}_{\mathrm{eff}}^{(\text{l})})(\mathbf{S}_{\mathrm{eff}}^{(\text{m})}\cdot\mathbf{S}^{(\text{t})}),
\label{SubeV_HP}
\end{align}
where $J$ and $K$ denotes the bilinear and biquadratic couplings between the fractionalized effective $\frac{1}{2}$-spins of Ni$^{2+}$ ions, $\mathbf{S}_{\mathrm{eff}}$, and the trimer composite spin, $\mathbf{S}^{\mathrm{(t)}}$.

While the effective ionic spins $\mathbf{S}_{\mathrm{eff}}$ have negligible direct couplings to each other\,\cite{supplementary}, as in the bilayer nickelates\,\cite{Jiang2024}, in La$_4$Ni$_3$O$_{10}$ they do couple to the interlayer $\mathbf{S}^{\mathrm{(t)}}$ as a result of $\tilde{J}_H$ in Eq.\,(\ref{SubeV_HP}).
These additional couplings are particularly strong for Ni$^{2+}$ ions in the outer (u and l) layers, producing an rather unexpected situation where the magnetic correlation between the disjoint outer layers are much stronger than their correlation to the adjacent middle layers.

\begin{figure}
\centering
\includegraphics[width=1\columnwidth]{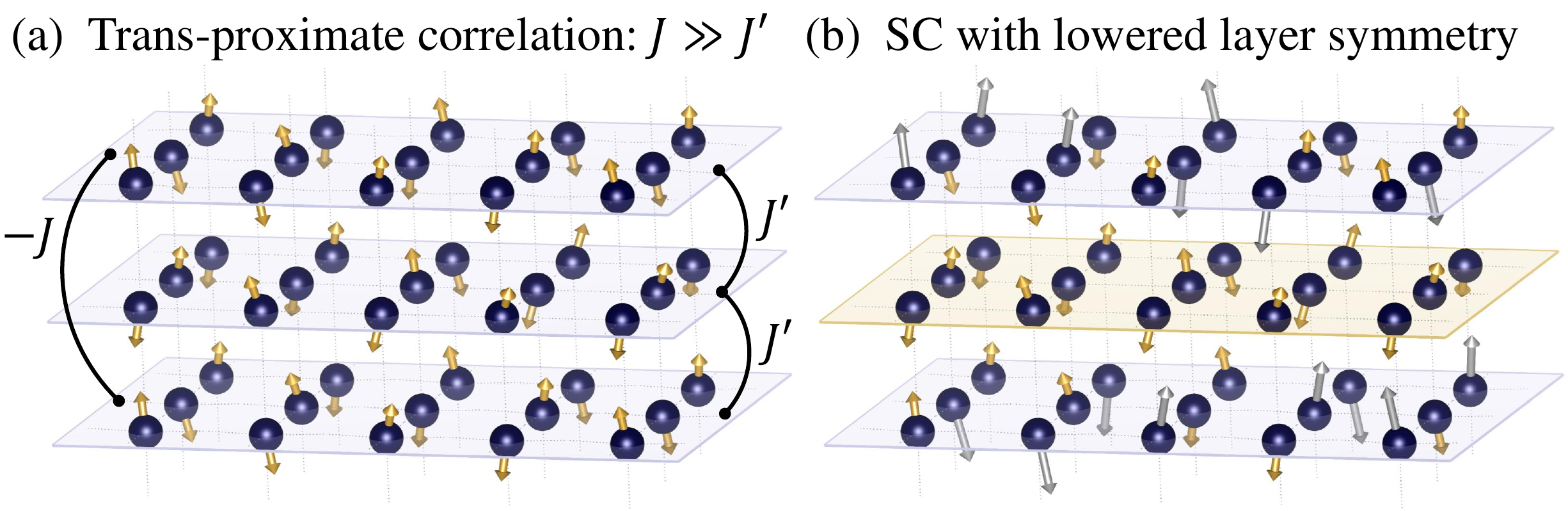}
\vspace{-0.8cm}
\caption{\textit{Illustration of the unusual `trans-proximate' correlation and possible enhancement of superconductivity with lowered local layer symmetry}
(a) Unusual trans-proximate interlayer correlations includes a strong correlation between disjoint outer layers that nonetheless correlate weakly with the adjacent middle layer.
(b) Possible enhancement of superconductivity (SC) via lowered local layer symmetry that allow more freely fluctuating $\frac{1}{2}$-spins connected throughout the sample.}
\vspace{-0.6cm}
\label{fig3}
\end{figure}

Such exotic `trans-proximate' interlayer correlation [c.f. Fig.\,\ref{fig3}(a)] between the disjoint outer layers is truly exceptional, since typical correlations, that emerge from combination of kinetic and interacting processes of shorter space-time scales~\cite{Anderson1950}, are always stronger at shorter range.
(Even in special cases with artificially designed ``topological order'', in which the short-range correlation can be completely suppressed~\cite{kitaev2006}, one still cannot find a stronger longer-range correlation.)
In layered materials, one thus expects stronger correlations between adjacent layers, through which weaker correlations are established between disjoint layers.
In pressurized La$_4$Ni$_3$O$_{10}$, the situation is quite distinct, since the interlayer correlations are mainly through couplings to the cross-layer trimer $\frac{1}{2}$-spin.
However, the trimer contains a strong quantum fluctuation that weakens its coupling to the middle layer and thus produces the exotic trans-proximate correlation between the disjoint outer layers.

This unusual trans-proximate spin correlation has an immediate physical consequence, namely a weaker superconducting phase stiffness in the outer layers, as a result of suppressed low-energy spin fluctuations that facilitate the kinetic processes of carriers.
This effect offers perhaps the simplest explanation for the lower superconducting transition temperature in La$_4$Ni$_3$O$_{10}$\,\cite{Zhu2024Superconductivity} compared to La$_3$Ni$_2$O$_7$, as only the middle layer experiences a weaker interlayer spin correlation to allow stiffer superconductivity, as illustrated in Fig.\,\ref{fig3}(a).
Furthermore, impurities with the La$_4$Ni$_3$O$_{10}$ structure have recently been identified in La$_3$Ni$_2$O$_7$ samples as the primary cause for the observed low superconducting volume fraction\,\cite{Wang2024Bulk, Li2025Ambient}, again reflecting the weaker superconductivity in the trilayer structure.

The above consideration suggests that the superconducting properties of La$_4$Ni$_3$O$_{10}$ could be significantly enhanced by lowering the trilayer symmetry to promote dimerization of only two of the Ni$^{2+}$ ions in the trilayer.
As an illustration, we consider a particular site Ni-v-Ni-O-Ni with an oxygen vacancy v in the trilayer structure that effectively decouples the Ni$^\mathrm{(u)}$ ion from the Ni$^\mathrm{(m)}$ and Ni$^\mathrm{(l)}$ ions\,\cite{supplementary}, or equivalently removes the $J_{ZZ}$ term in $H^\mathrm{(eV)}$ that couples Ni$^\mathrm{(u)}$ and Ni$^\mathrm{(m)}$.
As shown at the bottom of Fig.\,\ref{fig2}, the sub-eV dynamics would then resemble those of the bilayer nickelate\,\cite{Jiang2024} with an additional spin-1 Ni$^\mathrm{(u)}$ as a bystander.
Even though this would leave behind an unfavorable spin-1 Ni$^\mathrm{(u)}$ in the upper layer, the recovery of nearly uncoupled spin-$\frac{1}{2}$ Ni$^\mathrm{(m)}$ and Ni$^\mathrm{(l)}$ ions in the middle and lower layers (c.f. Tab.\,\ref{tab3} and Fig.\,\ref{fig3}(b)) would lead to efficient spin-fluctuations to facilitate kinetic processes of itinerant carriers and in turn stiffen the superconducting phase.

Interestingly, the recently identified superconducting La$_5$Ni$_3$O$_{11}$\,\cite{Mengzhu2025Superconductivity} appears to be a realization of the above scenario: by splitting the trilayer into a bilayer+monolayer superlattice, this material coherently lowers the trilayer symmetry as proposed above.
The observed enhanced transition temperature at around $64$\,K\,\cite{Mengzhu2025Superconductivity}, is therefore perfectly expected from the above analysis.

The enhancement of superconductivity via reducing trilayer symmetry suggests a further interesting possibility: in La$_4$Ni$_3$O$_{10}$ samples that host superconductivity, their local structures might already contain a lower trilayer symmetry.
In fact, the reported O vacancy in La$_4$Ni$_3$O$_{10}$\,\cite{Zhu2024Superconductivity, Yatoo2023Neutron} precisely lowers the local trilayer symmetry.
Furthermore, in the monolayer-trilayer superlattice of La$_3$Ni$_2$O$_7$ the observed strong fluctuations in apical O-Ni bond lengths\,\cite{Puphal2024Unconventional} also suggest a lower trilayer symmetry in the high-pressure phase.
It would be interesting to further investigate such lower trilayer symmetry via, for example, enhanced Debye-Waller factor (B-factor) of structural refinement, pair distribution function analysis of diffraction, nuclear quadrupole resonance, X-ray absorption fine structure, or an enhanced normal-state dielectric response.

In summary, we report an exotic pressure-induced trans-proximate correlation across layers (i.e. stronger correlation between disjoint layers than the adjacent ones) accompanying the emergence of a cross-layer trimer in La$_4$Ni$_3$O$_{10}$.
This rare correlation naturally explains the weaker superconductivity in this material by limiting the fluctuation of the fractionalized $\frac{1}{2}$-spin of Ni$^{2+}$ ions.
Our finding suggests a route for enhancing superconductivity in trilayer nickelates through lowering their layer symmetries and thus unifying the superconducting mechanism with previous nickelate and cuprate superconductors.
In general, such pressure-induced trans-proximate correlation is expected in many functional materials to produce rich classes of uncharted quantum states of matter.

This work is supported by the National Natural Science Foundation of China (NSFC) under Grant Nos. 12274287 and 12042507, and by the Innovation Program for Quantum Science and Technology No. 2021ZD0301900.
R.J. and B.M. are supported by a UKRI Future Leaders Fellowship [MR/V023926/1]. 
B.M. also acknowledges support from the Gianna Angelopoulos Programme for Science, Technology, and Innovation.

\bibliography{main.bib}

\begin{thebibliography}{95}%
\makeatletter
\providecommand \@ifxundefined [1]{%
 \@ifx{#1\undefined}
}%
\providecommand \@ifnum [1]{%
 \ifnum #1\expandafter \@firstoftwo
 \else \expandafter \@secondoftwo
 \fi
}%
\providecommand \@ifx [1]{%
 \ifx #1\expandafter \@firstoftwo
 \else \expandafter \@secondoftwo
 \fi
}%
\providecommand \natexlab [1]{#1}%
\providecommand \enquote  [1]{``#1''}%
\providecommand \bibnamefont  [1]{#1}%
\providecommand \bibfnamefont [1]{#1}%
\providecommand \citenamefont [1]{#1}%
\providecommand \href@noop [0]{\@secondoftwo}%
\providecommand \href [0]{\begingroup \@sanitize@url \@href}%
\providecommand \@href[1]{\@@startlink{#1}\@@href}%
\providecommand \@@href[1]{\endgroup#1\@@endlink}%
\providecommand \@sanitize@url [0]{\catcode `\\12\catcode `\$12\catcode `\&12\catcode `\#12\catcode `\^12\catcode `\_12\catcode `\%12\relax}%
\providecommand \@@startlink[1]{}%
\providecommand \@@endlink[0]{}%
\providecommand \url  [0]{\begingroup\@sanitize@url \@url }%
\providecommand \@url [1]{\endgroup\@href {#1}{\urlprefix }}%
\providecommand \urlprefix  [0]{URL }%
\providecommand \Eprint [0]{\href }%
\providecommand \doibase [0]{http://dx.doi.org/}%
\providecommand \selectlanguage [0]{\@gobble}%
\providecommand \bibinfo  [0]{\@secondoftwo}%
\providecommand \bibfield  [0]{\@secondoftwo}%
\providecommand \translation [1]{[#1]}%
\providecommand \BibitemOpen [0]{}%
\providecommand \bibitemStop [0]{}%
\providecommand \bibitemNoStop [0]{.\EOS\space}%
\providecommand \EOS [0]{\spacefactor3000\relax}%
\providecommand \BibitemShut  [1]{\csname bibitem#1\endcsname}%
\let\auto@bib@innerbib\@empty
\bibitem [{\citenamefont {Dagotto}(2005)}]{Dagotto2005}%
  \BibitemOpen
  \bibfield  {author} {\bibinfo {author} {\bibfnamefont {E.}~\bibnamefont {Dagotto}},\ }\href {\doibase 10.1126/science.1107559} {\bibfield  {journal} {\bibinfo  {journal} {Science}\ }\textbf {\bibinfo {volume} {309}},\ \bibinfo {pages} {257} (\bibinfo {year} {2005})}\BibitemShut {NoStop}%
\bibitem [{\citenamefont {Yang}\ \emph {et~al.}(2023)\citenamefont {Yang}, \citenamefont {Yu}, \citenamefont {Wen}, \citenamefont {Gui}, \citenamefont {Zhang}, \citenamefont {Zhan}, \citenamefont {Wang}, \citenamefont {Ying},\ and\ \citenamefont {Chen}}]{Yang2023Pressure-induced}%
  \BibitemOpen
  \bibfield  {author} {\bibinfo {author} {\bibfnamefont {Y.}~\bibnamefont {Yang}}, \bibinfo {author} {\bibfnamefont {F.}~\bibnamefont {Yu}}, \bibinfo {author} {\bibfnamefont {X.}~\bibnamefont {Wen}}, \bibinfo {author} {\bibfnamefont {Z.}~\bibnamefont {Gui}}, \bibinfo {author} {\bibfnamefont {Y.}~\bibnamefont {Zhang}}, \bibinfo {author} {\bibfnamefont {F.}~\bibnamefont {Zhan}}, \bibinfo {author} {\bibfnamefont {R.}~\bibnamefont {Wang}}, \bibinfo {author} {\bibfnamefont {J.}~\bibnamefont {Ying}}, \ and\ \bibinfo {author} {\bibfnamefont {X.}~\bibnamefont {Chen}},\ }\href {\doibase 10.1038/s41467-023-37971-2} {\bibfield  {journal} {\bibinfo  {journal} {Nature Communications}\ }\textbf {\bibinfo {volume} {14}} (\bibinfo {year} {2023}),\ 10.1038/s41467-023-37971-2}\BibitemShut {NoStop}%
\bibitem [{\citenamefont {Han}\ \emph {et~al.}(2015)\citenamefont {Han}, \citenamefont {Zhou}, \citenamefont {Malliakas}, \citenamefont {Duxbury}, \citenamefont {Mahanti}, \citenamefont {Kanatzidis},\ and\ \citenamefont {Ruan}}]{Tzong-Ru2015Exploration}%
  \BibitemOpen
  \bibfield  {author} {\bibinfo {author} {\bibfnamefont {T.-R.~T.}\ \bibnamefont {Han}}, \bibinfo {author} {\bibfnamefont {F.}~\bibnamefont {Zhou}}, \bibinfo {author} {\bibfnamefont {C.~D.}\ \bibnamefont {Malliakas}}, \bibinfo {author} {\bibfnamefont {P.~M.}\ \bibnamefont {Duxbury}}, \bibinfo {author} {\bibfnamefont {S.~D.}\ \bibnamefont {Mahanti}}, \bibinfo {author} {\bibfnamefont {M.~G.}\ \bibnamefont {Kanatzidis}}, \ and\ \bibinfo {author} {\bibfnamefont {C.-Y.}\ \bibnamefont {Ruan}},\ }\href {\doibase 10.1126/sciadv.1400173} {\bibfield  {journal} {\bibinfo  {journal} {Science Advances}\ }\textbf {\bibinfo {volume} {1}},\ \bibinfo {pages} {e1400173} (\bibinfo {year} {2015})},\ \Eprint {http://arxiv.org/abs/https://www.science.org/doi/pdf/10.1126/sciadv.1400173} {https://www.science.org/doi/pdf/10.1126/sciadv.1400173} \BibitemShut {NoStop}%
\bibitem [{\citenamefont {Sheng}\ and\ \citenamefont {Cooper}(1996)}]{Sheng1996Pressure-induced}%
  \BibitemOpen
  \bibfield  {author} {\bibinfo {author} {\bibfnamefont {Q.}~\bibnamefont {Sheng}}\ and\ \bibinfo {author} {\bibfnamefont {B.~R.}\ \bibnamefont {Cooper}},\ }\href {\doibase https://doi.org/10.1016/S0304-8853(96)00409-X} {\bibfield  {journal} {\bibinfo  {journal} {Journal of Magnetism and Magnetic Materials}\ }\textbf {\bibinfo {volume} {164}},\ \bibinfo {pages} {335} (\bibinfo {year} {1996})}\BibitemShut {NoStop}%
\bibitem [{\citenamefont {{Zhou}}\ \emph {et~al.}(2022)\citenamefont {{Zhou}}, \citenamefont {{Guo}}, \citenamefont {{Cai}}, \citenamefont {{Zhao}}, \citenamefont {{Gu}}, \citenamefont {{Lin}}, \citenamefont {{Yan}}, \citenamefont {{Huang}}, \citenamefont {{Yang}}, \citenamefont {{Long}}, \citenamefont {{Gong}}, \citenamefont {{Li}}, \citenamefont {{Li}}, \citenamefont {{Wu}}, \citenamefont {{Hu}}, \citenamefont {{Zhou}}, \citenamefont {{Xiang}},\ and\ \citenamefont {{Sun}}}]{Zhou2022Quantum}%
  \BibitemOpen
  \bibfield  {author} {\bibinfo {author} {\bibfnamefont {Y.}~\bibnamefont {{Zhou}}}, \bibinfo {author} {\bibfnamefont {J.}~\bibnamefont {{Guo}}}, \bibinfo {author} {\bibfnamefont {S.}~\bibnamefont {{Cai}}}, \bibinfo {author} {\bibfnamefont {J.}~\bibnamefont {{Zhao}}}, \bibinfo {author} {\bibfnamefont {G.}~\bibnamefont {{Gu}}}, \bibinfo {author} {\bibfnamefont {C.}~\bibnamefont {{Lin}}}, \bibinfo {author} {\bibfnamefont {H.}~\bibnamefont {{Yan}}}, \bibinfo {author} {\bibfnamefont {C.}~\bibnamefont {{Huang}}}, \bibinfo {author} {\bibfnamefont {C.}~\bibnamefont {{Yang}}}, \bibinfo {author} {\bibfnamefont {S.}~\bibnamefont {{Long}}}, \bibinfo {author} {\bibfnamefont {Y.}~\bibnamefont {{Gong}}}, \bibinfo {author} {\bibfnamefont {Y.}~\bibnamefont {{Li}}}, \bibinfo {author} {\bibfnamefont {X.}~\bibnamefont {{Li}}}, \bibinfo {author} {\bibfnamefont {Q.}~\bibnamefont {{Wu}}}, \bibinfo {author} {\bibfnamefont {J.}~\bibnamefont {{Hu}}}, \bibinfo {author} {\bibfnamefont {X.}~\bibnamefont {{Zhou}}}, \bibinfo {author}
  {\bibfnamefont {T.}~\bibnamefont {{Xiang}}}, \ and\ \bibinfo {author} {\bibfnamefont {L.}~\bibnamefont {{Sun}}},\ }\href {\doibase 10.1038/s41567-022-01513-2} {\bibfield  {journal} {\bibinfo  {journal} {Nature Physics}\ }\textbf {\bibinfo {volume} {18}},\ \bibinfo {pages} {406} (\bibinfo {year} {2022})}\BibitemShut {NoStop}%
\bibitem [{\citenamefont {Zapf}\ and\ \citenamefont {Dressel}(2016)}]{Zapf2016Europium}%
  \BibitemOpen
  \bibfield  {author} {\bibinfo {author} {\bibfnamefont {S.}~\bibnamefont {Zapf}}\ and\ \bibinfo {author} {\bibfnamefont {M.}~\bibnamefont {Dressel}},\ }\href {\doibase 10.1088/0034-4885/80/1/016501} {\bibfield  {journal} {\bibinfo  {journal} {Reports on progress in physics. Physical Society (Great Britain)}\ }\textbf {\bibinfo {volume} {80}},\ \bibinfo {pages} {016501} (\bibinfo {year} {2016})}\BibitemShut {NoStop}%
\bibitem [{\citenamefont {Sun}\ \emph {et~al.}(2015)\citenamefont {Sun}, \citenamefont {Matsuura}, \citenamefont {Ye}, \citenamefont {Mizukami}, \citenamefont {Shimozawa}, \citenamefont {Matsubayashi}, \citenamefont {Yamashita}, \citenamefont {Watashige}, \citenamefont {Kasahara}, \citenamefont {Matsuda}, \citenamefont {Yan}, \citenamefont {Sales}, \citenamefont {Uwatoko}, \citenamefont {Cheng},\ and\ \citenamefont {Shibauchi}}]{Sun2015Dome-shaped}%
  \BibitemOpen
  \bibfield  {author} {\bibinfo {author} {\bibfnamefont {J.}~\bibnamefont {Sun}}, \bibinfo {author} {\bibfnamefont {K.}~\bibnamefont {Matsuura}}, \bibinfo {author} {\bibfnamefont {G.}~\bibnamefont {Ye}}, \bibinfo {author} {\bibfnamefont {Y.}~\bibnamefont {Mizukami}}, \bibinfo {author} {\bibfnamefont {M.}~\bibnamefont {Shimozawa}}, \bibinfo {author} {\bibfnamefont {K.}~\bibnamefont {Matsubayashi}}, \bibinfo {author} {\bibfnamefont {M.}~\bibnamefont {Yamashita}}, \bibinfo {author} {\bibfnamefont {T.}~\bibnamefont {Watashige}}, \bibinfo {author} {\bibfnamefont {S.}~\bibnamefont {Kasahara}}, \bibinfo {author} {\bibfnamefont {Y.}~\bibnamefont {Matsuda}}, \bibinfo {author} {\bibfnamefont {J.-Q.}\ \bibnamefont {Yan}}, \bibinfo {author} {\bibfnamefont {B.}~\bibnamefont {Sales}}, \bibinfo {author} {\bibfnamefont {Y.}~\bibnamefont {Uwatoko}}, \bibinfo {author} {\bibfnamefont {J.-G.}\ \bibnamefont {Cheng}}, \ and\ \bibinfo {author} {\bibfnamefont {T.}~\bibnamefont {Shibauchi}},\ }\href {\doibase 10.1038/ncomms12146}
  {\bibfield  {journal} {\bibinfo  {journal} {Nature Communications}\ }\textbf {\bibinfo {volume} {7}} (\bibinfo {year} {2015}),\ 10.1038/ncomms12146}\BibitemShut {NoStop}%
\bibitem [{\citenamefont {Reiss}\ \emph {et~al.}(2020)\citenamefont {Reiss}, \citenamefont {Graf}, \citenamefont {Haghighirad}, \citenamefont {Knafo}, \citenamefont {Drigo}, \citenamefont {Bristow}, \citenamefont {Schofield},\ and\ \citenamefont {Coldea}}]{Reiss2020Quenched}%
  \BibitemOpen
  \bibfield  {author} {\bibinfo {author} {\bibfnamefont {P.}~\bibnamefont {Reiss}}, \bibinfo {author} {\bibfnamefont {D.}~\bibnamefont {Graf}}, \bibinfo {author} {\bibfnamefont {A.}~\bibnamefont {Haghighirad}}, \bibinfo {author} {\bibfnamefont {W.}~\bibnamefont {Knafo}}, \bibinfo {author} {\bibfnamefont {L.}~\bibnamefont {Drigo}}, \bibinfo {author} {\bibfnamefont {M.}~\bibnamefont {Bristow}}, \bibinfo {author} {\bibfnamefont {A.}~\bibnamefont {Schofield}}, \ and\ \bibinfo {author} {\bibfnamefont {A.}~\bibnamefont {Coldea}},\ }\href {\doibase 10.1038/s41567-019-0694-2} {\bibfield  {journal} {\bibinfo  {journal} {Nature Physics}\ }\textbf {\bibinfo {volume} {16}} (\bibinfo {year} {2020}),\ 10.1038/s41567-019-0694-2}\BibitemShut {NoStop}%
\bibitem [{\citenamefont {Wang}\ \emph {et~al.}(2022)\citenamefont {Wang}, \citenamefont {Yang}, \citenamefont {Yang}, \citenamefont {Chen}, \citenamefont {Zhang}, \citenamefont {Zhang}, \citenamefont {Zhu}, \citenamefont {Uwatoko}, \citenamefont {Gu}, \citenamefont {Dong}, \citenamefont {Sun}, \citenamefont {Jin},\ and\ \citenamefont {Cheng}}]{Wang2022Pressure-induced}%
  \BibitemOpen
  \bibfield  {author} {\bibinfo {author} {\bibfnamefont {N.~N.}\ \bibnamefont {Wang}}, \bibinfo {author} {\bibfnamefont {M.~W.}\ \bibnamefont {Yang}}, \bibinfo {author} {\bibfnamefont {Z.}~\bibnamefont {Yang}}, \bibinfo {author} {\bibfnamefont {K.~Y.}\ \bibnamefont {Chen}}, \bibinfo {author} {\bibfnamefont {H.}~\bibnamefont {Zhang}}, \bibinfo {author} {\bibfnamefont {Q.~H.}\ \bibnamefont {Zhang}}, \bibinfo {author} {\bibfnamefont {Z.~H.}\ \bibnamefont {Zhu}}, \bibinfo {author} {\bibfnamefont {Y.}~\bibnamefont {Uwatoko}}, \bibinfo {author} {\bibfnamefont {L.}~\bibnamefont {Gu}}, \bibinfo {author} {\bibfnamefont {X.~L.}\ \bibnamefont {Dong}}, \bibinfo {author} {\bibfnamefont {J.~P.}\ \bibnamefont {Sun}}, \bibinfo {author} {\bibfnamefont {K.~J.}\ \bibnamefont {Jin}}, \ and\ \bibinfo {author} {\bibfnamefont {J.-G.}\ \bibnamefont {Cheng}},\ }\href {\doibase 10.1038/s41467-022-32065-x} {\bibfield  {journal} {\bibinfo  {journal} {Nature Communications}\ }\textbf {\bibinfo {volume} {13}} (\bibinfo {year} {2022}),\
  10.1038/s41467-022-32065-x}\BibitemShut {NoStop}%
\bibitem [{\citenamefont {Dissanayake}\ \emph {et~al.}(2023)\citenamefont {Dissanayake}, \citenamefont {Matsuda}, \citenamefont {Yoshimi}, \citenamefont {Kasamatsu}, \citenamefont {Ye}, \citenamefont {Chi}, \citenamefont {Steinhardt}, \citenamefont {Fabbris}, \citenamefont {Haravifard}, \citenamefont {Cheng}, \citenamefont {Yan}, \citenamefont {Gouchi},\ and\ \citenamefont {Uwatoko}}]{Dissanayake2023Helical}%
  \BibitemOpen
  \bibfield  {author} {\bibinfo {author} {\bibfnamefont {S.}~\bibnamefont {Dissanayake}}, \bibinfo {author} {\bibfnamefont {M.}~\bibnamefont {Matsuda}}, \bibinfo {author} {\bibfnamefont {K.}~\bibnamefont {Yoshimi}}, \bibinfo {author} {\bibfnamefont {S.}~\bibnamefont {Kasamatsu}}, \bibinfo {author} {\bibfnamefont {F.}~\bibnamefont {Ye}}, \bibinfo {author} {\bibfnamefont {S.}~\bibnamefont {Chi}}, \bibinfo {author} {\bibfnamefont {W.}~\bibnamefont {Steinhardt}}, \bibinfo {author} {\bibfnamefont {G.}~\bibnamefont {Fabbris}}, \bibinfo {author} {\bibfnamefont {S.}~\bibnamefont {Haravifard}}, \bibinfo {author} {\bibfnamefont {J.}~\bibnamefont {Cheng}}, \bibinfo {author} {\bibfnamefont {J.-Q.}\ \bibnamefont {Yan}}, \bibinfo {author} {\bibfnamefont {J.}~\bibnamefont {Gouchi}}, \ and\ \bibinfo {author} {\bibfnamefont {Y.}~\bibnamefont {Uwatoko}},\ }\href {\doibase 10.1103/PhysRevResearch.5.043026} {\bibfield  {journal} {\bibinfo  {journal} {Physical Review Research}\ }\textbf {\bibinfo {volume} {5}} (\bibinfo {year}
  {2023}),\ 10.1103/PhysRevResearch.5.043026}\BibitemShut {NoStop}%
\bibitem [{\citenamefont {Alireza}\ \emph {et~al.}(2010)\citenamefont {Alireza}, \citenamefont {Nakamura}, \citenamefont {Goh}, \citenamefont {Maeno}, \citenamefont {Nakatsuji}, \citenamefont {Ko}, \citenamefont {Sutherland}, \citenamefont {Julian},\ and\ \citenamefont {Lonzarich}}]{Alireza2010Evidence}%
  \BibitemOpen
  \bibfield  {author} {\bibinfo {author} {\bibfnamefont {P.}~\bibnamefont {Alireza}}, \bibinfo {author} {\bibfnamefont {F.}~\bibnamefont {Nakamura}}, \bibinfo {author} {\bibfnamefont {S.}~\bibnamefont {Goh}}, \bibinfo {author} {\bibfnamefont {Y.}~\bibnamefont {Maeno}}, \bibinfo {author} {\bibfnamefont {S.}~\bibnamefont {Nakatsuji}}, \bibinfo {author} {\bibfnamefont {Y.}~\bibnamefont {Ko}}, \bibinfo {author} {\bibfnamefont {M.}~\bibnamefont {Sutherland}}, \bibinfo {author} {\bibfnamefont {S.}~\bibnamefont {Julian}}, \ and\ \bibinfo {author} {\bibfnamefont {G.}~\bibnamefont {Lonzarich}},\ }\href {\doibase 10.1088/0953-8984/22/5/052202} {\bibfield  {journal} {\bibinfo  {journal} {Journal of physics. Condensed matter : an Institute of Physics journal}\ }\textbf {\bibinfo {volume} {22}},\ \bibinfo {pages} {052202} (\bibinfo {year} {2010})}\BibitemShut {NoStop}%
\bibitem [{\citenamefont {Yu}\ \emph {et~al.}(2021)\citenamefont {Yu}, \citenamefont {Ma}, \citenamefont {Zhuo}, \citenamefont {Liu}, \citenamefont {Wen}, \citenamefont {Lei}, \citenamefont {Ying},\ and\ \citenamefont {Chen}}]{Yu2021Unusual}%
  \BibitemOpen
  \bibfield  {author} {\bibinfo {author} {\bibfnamefont {F.}~\bibnamefont {Yu}}, \bibinfo {author} {\bibfnamefont {D.}~\bibnamefont {Ma}}, \bibinfo {author} {\bibfnamefont {W.}~\bibnamefont {Zhuo}}, \bibinfo {author} {\bibfnamefont {S.}~\bibnamefont {Liu}}, \bibinfo {author} {\bibfnamefont {X.}~\bibnamefont {Wen}}, \bibinfo {author} {\bibfnamefont {B.}~\bibnamefont {Lei}}, \bibinfo {author} {\bibfnamefont {J.}~\bibnamefont {Ying}}, \ and\ \bibinfo {author} {\bibfnamefont {X.}~\bibnamefont {Chen}},\ }\href {\doibase 10.1038/s41467-021-23928-w} {\bibfield  {journal} {\bibinfo  {journal} {Nature Communications}\ }\textbf {\bibinfo {volume} {12}},\ \bibinfo {pages} {3645} (\bibinfo {year} {2021})}\BibitemShut {NoStop}%
\bibitem [{\citenamefont {Du}\ \emph {et~al.}(2021)\citenamefont {Du}, \citenamefont {Luo}, \citenamefont {Ortiz}, \citenamefont {Chen}, \citenamefont {Duan}, \citenamefont {Zhang}, \citenamefont {Lu}, \citenamefont {Wilson}, \citenamefont {Song},\ and\ \citenamefont {Yuan}}]{Du2021Pressure}%
  \BibitemOpen
  \bibfield  {author} {\bibinfo {author} {\bibfnamefont {F.}~\bibnamefont {Du}}, \bibinfo {author} {\bibfnamefont {S.}~\bibnamefont {Luo}}, \bibinfo {author} {\bibfnamefont {B.~R.}\ \bibnamefont {Ortiz}}, \bibinfo {author} {\bibfnamefont {Y.}~\bibnamefont {Chen}}, \bibinfo {author} {\bibfnamefont {W.}~\bibnamefont {Duan}}, \bibinfo {author} {\bibfnamefont {D.}~\bibnamefont {Zhang}}, \bibinfo {author} {\bibfnamefont {X.}~\bibnamefont {Lu}}, \bibinfo {author} {\bibfnamefont {S.~D.}\ \bibnamefont {Wilson}}, \bibinfo {author} {\bibfnamefont {Y.}~\bibnamefont {Song}}, \ and\ \bibinfo {author} {\bibfnamefont {H.}~\bibnamefont {Yuan}},\ }\href {\doibase 10.1103/PhysRevB.103.L220504} {\bibfield  {journal} {\bibinfo  {journal} {Phys. Rev. B}\ }\textbf {\bibinfo {volume} {103}},\ \bibinfo {pages} {L220504} (\bibinfo {year} {2021})}\BibitemShut {NoStop}%
\bibitem [{\citenamefont {Nie}\ \emph {et~al.}(2023)\citenamefont {Nie}, \citenamefont {Yang}, \citenamefont {Zhang}, \citenamefont {Liu}, \citenamefont {Xia}, \citenamefont {Dai}, \citenamefont {Zhao}, \citenamefont {Tu}, \citenamefont {Kong}, \citenamefont {Jin}, \citenamefont {Guo},\ and\ \citenamefont {Li}}]{Nie2023Pressure}%
  \BibitemOpen
  \bibfield  {author} {\bibinfo {author} {\bibfnamefont {J.}~\bibnamefont {Nie}}, \bibinfo {author} {\bibfnamefont {X.}~\bibnamefont {Yang}}, \bibinfo {author} {\bibfnamefont {X.}~\bibnamefont {Zhang}}, \bibinfo {author} {\bibfnamefont {X.}~\bibnamefont {Liu}}, \bibinfo {author} {\bibfnamefont {W.}~\bibnamefont {Xia}}, \bibinfo {author} {\bibfnamefont {D.}~\bibnamefont {Dai}}, \bibinfo {author} {\bibfnamefont {C.}~\bibnamefont {Zhao}}, \bibinfo {author} {\bibfnamefont {C.}~\bibnamefont {Tu}}, \bibinfo {author} {\bibfnamefont {X.}~\bibnamefont {Kong}}, \bibinfo {author} {\bibfnamefont {X.}~\bibnamefont {Jin}}, \bibinfo {author} {\bibfnamefont {Y.}~\bibnamefont {Guo}}, \ and\ \bibinfo {author} {\bibfnamefont {S.}~\bibnamefont {Li}},\ }\href@noop {} {\enquote {\bibinfo {title} {Pressure-induced double-dome superconductivity in kagome metal $\rm{ CsBi_3Ti_5}$},}\ } (\bibinfo {year} {2023}),\ \Eprint {http://arxiv.org/abs/2308.10129} {arXiv:2308.10129 [cond-mat.supr-con]} \BibitemShut {NoStop}%
\bibitem [{\citenamefont {Ran}\ \emph {et~al.}(2021)\citenamefont {Ran}, \citenamefont {Saha}, \citenamefont {Liu}, \citenamefont {Graf}, \citenamefont {Paglione},\ and\ \citenamefont {Butch}}]{Ran2021Expansion}%
  \BibitemOpen
  \bibfield  {author} {\bibinfo {author} {\bibfnamefont {S.}~\bibnamefont {Ran}}, \bibinfo {author} {\bibfnamefont {S.~R.}\ \bibnamefont {Saha}}, \bibinfo {author} {\bibfnamefont {I.-L.}\ \bibnamefont {Liu}}, \bibinfo {author} {\bibfnamefont {D.}~\bibnamefont {Graf}}, \bibinfo {author} {\bibfnamefont {J.}~\bibnamefont {Paglione}}, \ and\ \bibinfo {author} {\bibfnamefont {N.~P.}\ \bibnamefont {Butch}},\ }\href {\doibase 10.1038/s41535-021-00376-9} {\bibfield  {journal} {\bibinfo  {journal} {npj Quantum Materials}\ }\textbf {\bibinfo {volume} {6}} (\bibinfo {year} {2021}),\ 10.1038/s41535-021-00376-9}\BibitemShut {NoStop}%
\bibitem [{\citenamefont {Sun}\ \emph {et~al.}(2023)\citenamefont {Sun}, \citenamefont {Huo}, \citenamefont {Hu}, \citenamefont {Li}, \citenamefont {Liu}, \citenamefont {Han}, \citenamefont {Tang}, \citenamefont {Mao}, \citenamefont {Yang}, \citenamefont {Wang}, \citenamefont {Cheng}, \citenamefont {Yao}, \citenamefont {Zhang},\ and\ \citenamefont {Wang}}]{Hualei0516}%
  \BibitemOpen
  \bibfield  {author} {\bibinfo {author} {\bibfnamefont {H.}~\bibnamefont {Sun}}, \bibinfo {author} {\bibfnamefont {M.}~\bibnamefont {Huo}}, \bibinfo {author} {\bibfnamefont {X.}~\bibnamefont {Hu}}, \bibinfo {author} {\bibfnamefont {J.}~\bibnamefont {Li}}, \bibinfo {author} {\bibfnamefont {Z.}~\bibnamefont {Liu}}, \bibinfo {author} {\bibfnamefont {Y.}~\bibnamefont {Han}}, \bibinfo {author} {\bibfnamefont {L.}~\bibnamefont {Tang}}, \bibinfo {author} {\bibfnamefont {Z.}~\bibnamefont {Mao}}, \bibinfo {author} {\bibfnamefont {P.}~\bibnamefont {Yang}}, \bibinfo {author} {\bibfnamefont {B.}~\bibnamefont {Wang}}, \bibinfo {author} {\bibfnamefont {J.}~\bibnamefont {Cheng}}, \bibinfo {author} {\bibfnamefont {D.-X.}\ \bibnamefont {Yao}}, \bibinfo {author} {\bibfnamefont {G.-M.}\ \bibnamefont {Zhang}}, \ and\ \bibinfo {author} {\bibfnamefont {M.}~\bibnamefont {Wang}},\ }\href {\doibase 10.1038/s41586-023-06408-7} {\bibfield  {journal} {\bibinfo  {journal} {Nature}\ }\textbf {\bibinfo {volume} {621}},\ \bibinfo {pages}
  {493} (\bibinfo {year} {2023})}\BibitemShut {NoStop}%
\bibitem [{\citenamefont {Hou}\ \emph {et~al.}(2023)\citenamefont {Hou}, \citenamefont {Yang}, \citenamefont {Liu}, \citenamefont {Li}, \citenamefont {Shan}, \citenamefont {Ma}, \citenamefont {Wang}, \citenamefont {Wang}, \citenamefont {Guo}, \citenamefont {Sun}, \citenamefont {Uwatoko}, \citenamefont {Wang}, \citenamefont {Zhang}, \citenamefont {Wang},\ and\ \citenamefont {Cheng}}]{Hou0719}%
  \BibitemOpen
  \bibfield  {author} {\bibinfo {author} {\bibfnamefont {J.}~\bibnamefont {Hou}}, \bibinfo {author} {\bibfnamefont {P.}~\bibnamefont {Yang}}, \bibinfo {author} {\bibfnamefont {Z.-Y.}\ \bibnamefont {Liu}}, \bibinfo {author} {\bibfnamefont {J.-Y.}\ \bibnamefont {Li}}, \bibinfo {author} {\bibfnamefont {P.-F.}\ \bibnamefont {Shan}}, \bibinfo {author} {\bibfnamefont {L.}~\bibnamefont {Ma}}, \bibinfo {author} {\bibfnamefont {G.}~\bibnamefont {Wang}}, \bibinfo {author} {\bibfnamefont {N.}~\bibnamefont {Wang}}, \bibinfo {author} {\bibfnamefont {H.}~\bibnamefont {Guo}}, \bibinfo {author} {\bibfnamefont {J.}~\bibnamefont {Sun}}, \bibinfo {author} {\bibfnamefont {Y.}~\bibnamefont {Uwatoko}}, \bibinfo {author} {\bibfnamefont {M.}~\bibnamefont {Wang}}, \bibinfo {author} {\bibfnamefont {G.-M.}\ \bibnamefont {Zhang}}, \bibinfo {author} {\bibfnamefont {B.-S.}\ \bibnamefont {Wang}}, \ and\ \bibinfo {author} {\bibfnamefont {J.-G.}\ \bibnamefont {Cheng}},\ }\href
  {https://iopscience.iop.org/article/10.1088/0256-307X/40/11/117302} {\bibfield  {journal} {\bibinfo  {journal} {Chinese Physics Letters}\ }\textbf {\bibinfo {volume} {40}},\ \bibinfo {pages} {117302} (\bibinfo {year} {2023})}\BibitemShut {NoStop}%
\bibitem [{\citenamefont {Zhang}\ \emph {et~al.}(2024{\natexlab{a}})\citenamefont {Zhang}, \citenamefont {Su}, \citenamefont {Huang}, \citenamefont {Shan}, \citenamefont {Sun}, \citenamefont {Huo}, \citenamefont {Ye}, \citenamefont {Zhang}, \citenamefont {Yang}, \citenamefont {Xu}, \citenamefont {Su}, \citenamefont {Li}, \citenamefont {Smidman}, \citenamefont {Wang}, \citenamefont {Jiao},\ and\ \citenamefont {Yuan}}]{Yanan0727}%
  \BibitemOpen
  \bibfield  {author} {\bibinfo {author} {\bibfnamefont {Y.}~\bibnamefont {Zhang}}, \bibinfo {author} {\bibfnamefont {D.}~\bibnamefont {Su}}, \bibinfo {author} {\bibfnamefont {Y.}~\bibnamefont {Huang}}, \bibinfo {author} {\bibfnamefont {Z.}~\bibnamefont {Shan}}, \bibinfo {author} {\bibfnamefont {H.}~\bibnamefont {Sun}}, \bibinfo {author} {\bibfnamefont {M.}~\bibnamefont {Huo}}, \bibinfo {author} {\bibfnamefont {K.}~\bibnamefont {Ye}}, \bibinfo {author} {\bibfnamefont {J.}~\bibnamefont {Zhang}}, \bibinfo {author} {\bibfnamefont {Z.}~\bibnamefont {Yang}}, \bibinfo {author} {\bibfnamefont {Y.}~\bibnamefont {Xu}}, \bibinfo {author} {\bibfnamefont {Y.}~\bibnamefont {Su}}, \bibinfo {author} {\bibfnamefont {R.}~\bibnamefont {Li}}, \bibinfo {author} {\bibfnamefont {M.}~\bibnamefont {Smidman}}, \bibinfo {author} {\bibfnamefont {M.}~\bibnamefont {Wang}}, \bibinfo {author} {\bibfnamefont {L.}~\bibnamefont {Jiao}}, \ and\ \bibinfo {author} {\bibfnamefont {H.}~\bibnamefont {Yuan}},\ }\href {\doibase
  10.1038/s41567-024-02515-y} {\bibfield  {journal} {\bibinfo  {journal} {Nature Physics}\ }\textbf {\bibinfo {volume} {20}},\ \bibinfo {pages} {1269} (\bibinfo {year} {2024}{\natexlab{a}})}\BibitemShut {NoStop}%
\bibitem [{\citenamefont {Goodge}\ \emph {et~al.}(2021)\citenamefont {Goodge}, \citenamefont {Li}, \citenamefont {Lee}, \citenamefont {Osada}, \citenamefont {Wang}, \citenamefont {Sawatzky}, \citenamefont {Hwang},\ and\ \citenamefont {Kourkoutis}}]{Berit2021Doping}%
  \BibitemOpen
  \bibfield  {author} {\bibinfo {author} {\bibfnamefont {B.~H.}\ \bibnamefont {Goodge}}, \bibinfo {author} {\bibfnamefont {D.}~\bibnamefont {Li}}, \bibinfo {author} {\bibfnamefont {K.}~\bibnamefont {Lee}}, \bibinfo {author} {\bibfnamefont {M.}~\bibnamefont {Osada}}, \bibinfo {author} {\bibfnamefont {B.~Y.}\ \bibnamefont {Wang}}, \bibinfo {author} {\bibfnamefont {G.~A.}\ \bibnamefont {Sawatzky}}, \bibinfo {author} {\bibfnamefont {H.~Y.}\ \bibnamefont {Hwang}}, \ and\ \bibinfo {author} {\bibfnamefont {L.~F.}\ \bibnamefont {Kourkoutis}},\ }\href {\doibase 10.1073/pnas.2007683118} {\bibfield  {journal} {\bibinfo  {journal} {Proceedings of the National Academy of Sciences}\ }\textbf {\bibinfo {volume} {118}},\ \bibinfo {pages} {e2007683118} (\bibinfo {year} {2021})}\BibitemShut {NoStop}%
\bibitem [{\citenamefont {Ding}\ \emph {et~al.}(2024)\citenamefont {Ding}, \citenamefont {Fan}, \citenamefont {Wang}, \citenamefont {Li}, \citenamefont {An}, \citenamefont {Ye}, \citenamefont {Tang}, \citenamefont {Lei}, \citenamefont {Sun}, \citenamefont {Guo}, \citenamefont {Chen}, \citenamefont {Sangphet}, \citenamefont {Wang}, \citenamefont {Xu}, \citenamefont {Peng},\ and\ \citenamefont {Feng}}]{Ding2024Cuprate-like}%
  \BibitemOpen
  \bibfield  {author} {\bibinfo {author} {\bibfnamefont {X.}~\bibnamefont {Ding}}, \bibinfo {author} {\bibfnamefont {Y.}~\bibnamefont {Fan}}, \bibinfo {author} {\bibfnamefont {X.}~\bibnamefont {Wang}}, \bibinfo {author} {\bibfnamefont {C.}~\bibnamefont {Li}}, \bibinfo {author} {\bibfnamefont {Z.}~\bibnamefont {An}}, \bibinfo {author} {\bibfnamefont {J.}~\bibnamefont {Ye}}, \bibinfo {author} {\bibfnamefont {S.}~\bibnamefont {Tang}}, \bibinfo {author} {\bibfnamefont {M.}~\bibnamefont {Lei}}, \bibinfo {author} {\bibfnamefont {X.}~\bibnamefont {Sun}}, \bibinfo {author} {\bibfnamefont {N.}~\bibnamefont {Guo}}, \bibinfo {author} {\bibfnamefont {Z.}~\bibnamefont {Chen}}, \bibinfo {author} {\bibfnamefont {S.}~\bibnamefont {Sangphet}}, \bibinfo {author} {\bibfnamefont {Y.}~\bibnamefont {Wang}}, \bibinfo {author} {\bibfnamefont {H.}~\bibnamefont {Xu}}, \bibinfo {author} {\bibfnamefont {R.}~\bibnamefont {Peng}}, \ and\ \bibinfo {author} {\bibfnamefont {D.}~\bibnamefont {Feng}},\ }\href {\doibase 10.1093/nsr/nwae194}
  {\bibfield  {journal} {\bibinfo  {journal} {National Science Review}\ }\textbf {\bibinfo {volume} {11}},\ \bibinfo {pages} {nwae194} (\bibinfo {year} {2024})}\BibitemShut {NoStop}%
\bibitem [{\citenamefont {Lang}\ \emph {et~al.}(2021)\citenamefont {Lang}, \citenamefont {Jiang},\ and\ \citenamefont {Ku}}]{lang2021}%
  \BibitemOpen
  \bibfield  {author} {\bibinfo {author} {\bibfnamefont {Z.-J.}\ \bibnamefont {Lang}}, \bibinfo {author} {\bibfnamefont {R.}~\bibnamefont {Jiang}}, \ and\ \bibinfo {author} {\bibfnamefont {W.}~\bibnamefont {Ku}},\ }\href {https://link.aps.org/doi/10.1103/PhysRevB.103.L180502} {\bibfield  {journal} {\bibinfo  {journal} {Phys. Rev. B}\ }\textbf {\bibinfo {volume} {103}},\ \bibinfo {pages} {L180502} (\bibinfo {year} {2021})}\BibitemShut {NoStop}%
\bibitem [{\citenamefont {Lang}\ \emph {et~al.}(2022)\citenamefont {Lang}, \citenamefont {Jiang},\ and\ \citenamefont {Ku}}]{lang2022}%
  \BibitemOpen
  \bibfield  {author} {\bibinfo {author} {\bibfnamefont {Z.-J.}\ \bibnamefont {Lang}}, \bibinfo {author} {\bibfnamefont {R.}~\bibnamefont {Jiang}}, \ and\ \bibinfo {author} {\bibfnamefont {W.}~\bibnamefont {Ku}},\ }\href {\doibase 10.1103/PhysRevB.105.L100501} {\bibfield  {journal} {\bibinfo  {journal} {Phys. Rev. B}\ }\textbf {\bibinfo {volume} {105}},\ \bibinfo {pages} {L100501} (\bibinfo {year} {2022})}\BibitemShut {NoStop}%
\bibitem [{\citenamefont {{Azuma}}\ \emph {et~al.}(1992)\citenamefont {{Azuma}}, \citenamefont {{Hiroi}}, \citenamefont {{Takano}}, \citenamefont {{Bando}},\ and\ \citenamefont {{Takeda}}}]{Azuma1992Superconductivity}%
  \BibitemOpen
  \bibfield  {author} {\bibinfo {author} {\bibfnamefont {M.}~\bibnamefont {{Azuma}}}, \bibinfo {author} {\bibfnamefont {Z.}~\bibnamefont {{Hiroi}}}, \bibinfo {author} {\bibfnamefont {M.}~\bibnamefont {{Takano}}}, \bibinfo {author} {\bibfnamefont {Y.}~\bibnamefont {{Bando}}}, \ and\ \bibinfo {author} {\bibfnamefont {Y.}~\bibnamefont {{Takeda}}},\ }\href {\doibase 10.1038/356775a0} {\bibfield  {journal} {\bibinfo  {journal} {Nature}\ }\textbf {\bibinfo {volume} {356}},\ \bibinfo {pages} {775} (\bibinfo {year} {1992})}\BibitemShut {NoStop}%
\bibitem [{\citenamefont {Jiang}\ \emph {et~al.}(2024{\natexlab{a}})\citenamefont {Jiang}, \citenamefont {Hou}, \citenamefont {Fan}, \citenamefont {Lang},\ and\ \citenamefont {Ku}}]{Jiang2024}%
  \BibitemOpen
  \bibfield  {author} {\bibinfo {author} {\bibfnamefont {R.}~\bibnamefont {Jiang}}, \bibinfo {author} {\bibfnamefont {J.}~\bibnamefont {Hou}}, \bibinfo {author} {\bibfnamefont {Z.}~\bibnamefont {Fan}}, \bibinfo {author} {\bibfnamefont {Z.-J.}\ \bibnamefont {Lang}}, \ and\ \bibinfo {author} {\bibfnamefont {W.}~\bibnamefont {Ku}},\ }\href {\doibase 10.1103/PhysRevLett.132.126503} {\bibfield  {journal} {\bibinfo  {journal} {Phys. Rev. Lett.}\ }\textbf {\bibinfo {volume} {132}},\ \bibinfo {pages} {126503} (\bibinfo {year} {2024}{\natexlab{a}})}\BibitemShut {NoStop}%
\bibitem [{\citenamefont {Sakakibara}\ \emph {et~al.}(2024)\citenamefont {Sakakibara}, \citenamefont {Ochi}, \citenamefont {Nagata}, \citenamefont {Ueki}, \citenamefont {Sakurai}, \citenamefont {Matsumoto}, \citenamefont {Terashima}, \citenamefont {Hirose}, \citenamefont {Ohta}, \citenamefont {Kato}, \citenamefont {Takano},\ and\ \citenamefont {Kuroki}}]{Sakakibara2024Theoretical}%
  \BibitemOpen
  \bibfield  {author} {\bibinfo {author} {\bibfnamefont {H.}~\bibnamefont {Sakakibara}}, \bibinfo {author} {\bibfnamefont {M.}~\bibnamefont {Ochi}}, \bibinfo {author} {\bibfnamefont {H.}~\bibnamefont {Nagata}}, \bibinfo {author} {\bibfnamefont {Y.}~\bibnamefont {Ueki}}, \bibinfo {author} {\bibfnamefont {H.}~\bibnamefont {Sakurai}}, \bibinfo {author} {\bibfnamefont {R.}~\bibnamefont {Matsumoto}}, \bibinfo {author} {\bibfnamefont {K.}~\bibnamefont {Terashima}}, \bibinfo {author} {\bibfnamefont {K.}~\bibnamefont {Hirose}}, \bibinfo {author} {\bibfnamefont {H.}~\bibnamefont {Ohta}}, \bibinfo {author} {\bibfnamefont {M.}~\bibnamefont {Kato}}, \bibinfo {author} {\bibfnamefont {Y.}~\bibnamefont {Takano}}, \ and\ \bibinfo {author} {\bibfnamefont {K.}~\bibnamefont {Kuroki}},\ }\href {\doibase 10.1103/PhysRevB.109.144511} {\bibfield  {journal} {\bibinfo  {journal} {Phys. Rev. B}\ }\textbf {\bibinfo {volume} {109}},\ \bibinfo {pages} {144511} (\bibinfo {year} {2024})}\BibitemShut {NoStop}%
\bibitem [{\citenamefont {Zhu}\ \emph {et~al.}(2024)\citenamefont {Zhu}, \citenamefont {Peng}, \citenamefont {Zhang}, \citenamefont {Pan}, \citenamefont {Chen}, \citenamefont {Chen}, \citenamefont {Ren}, \citenamefont {Liu}, \citenamefont {Hao}, \citenamefont {Li}, \citenamefont {Xing}, \citenamefont {Lan}, \citenamefont {Han}, \citenamefont {Wang}, \citenamefont {Jia}, \citenamefont {Wo}, \citenamefont {Gu}, \citenamefont {Gu}, \citenamefont {Ji}, \citenamefont {Wang}, \citenamefont {Gou}, \citenamefont {Shen}, \citenamefont {Ying}, \citenamefont {Chen}, \citenamefont {Yang}, \citenamefont {Cao}, \citenamefont {Zheng}, \citenamefont {Zeng}, \citenamefont {Guo},\ and\ \citenamefont {Zhao}}]{Zhu2024Superconductivity}%
  \BibitemOpen
  \bibfield  {author} {\bibinfo {author} {\bibfnamefont {Y.}~\bibnamefont {Zhu}}, \bibinfo {author} {\bibfnamefont {D.}~\bibnamefont {Peng}}, \bibinfo {author} {\bibfnamefont {E.}~\bibnamefont {Zhang}}, \bibinfo {author} {\bibfnamefont {B.}~\bibnamefont {Pan}}, \bibinfo {author} {\bibfnamefont {X.}~\bibnamefont {Chen}}, \bibinfo {author} {\bibfnamefont {L.}~\bibnamefont {Chen}}, \bibinfo {author} {\bibfnamefont {H.}~\bibnamefont {Ren}}, \bibinfo {author} {\bibfnamefont {F.}~\bibnamefont {Liu}}, \bibinfo {author} {\bibfnamefont {Y.}~\bibnamefont {Hao}}, \bibinfo {author} {\bibfnamefont {N.}~\bibnamefont {Li}}, \bibinfo {author} {\bibfnamefont {Z.}~\bibnamefont {Xing}}, \bibinfo {author} {\bibfnamefont {F.}~\bibnamefont {Lan}}, \bibinfo {author} {\bibfnamefont {J.}~\bibnamefont {Han}}, \bibinfo {author} {\bibfnamefont {J.}~\bibnamefont {Wang}}, \bibinfo {author} {\bibfnamefont {D.}~\bibnamefont {Jia}}, \bibinfo {author} {\bibfnamefont {H.}~\bibnamefont {Wo}}, \bibinfo {author} {\bibfnamefont {Y.}~\bibnamefont
  {Gu}}, \bibinfo {author} {\bibfnamefont {Y.}~\bibnamefont {Gu}}, \bibinfo {author} {\bibfnamefont {L.}~\bibnamefont {Ji}}, \bibinfo {author} {\bibfnamefont {W.}~\bibnamefont {Wang}}, \bibinfo {author} {\bibfnamefont {H.}~\bibnamefont {Gou}}, \bibinfo {author} {\bibfnamefont {Y.}~\bibnamefont {Shen}}, \bibinfo {author} {\bibfnamefont {T.}~\bibnamefont {Ying}}, \bibinfo {author} {\bibfnamefont {X.}~\bibnamefont {Chen}}, \bibinfo {author} {\bibfnamefont {W.}~\bibnamefont {Yang}}, \bibinfo {author} {\bibfnamefont {H.}~\bibnamefont {Cao}}, \bibinfo {author} {\bibfnamefont {C.}~\bibnamefont {Zheng}}, \bibinfo {author} {\bibfnamefont {Q.}~\bibnamefont {Zeng}}, \bibinfo {author} {\bibfnamefont {J.-g.}\ \bibnamefont {Guo}}, \ and\ \bibinfo {author} {\bibfnamefont {J.}~\bibnamefont {Zhao}},\ }\href {\doibase 10.1038/s41586-024-07553-3} {\bibfield  {journal} {\bibinfo  {journal} {Nature}\ }\textbf {\bibinfo {volume} {631}},\ \bibinfo {pages} {531} (\bibinfo {year} {2024})}\BibitemShut {NoStop}%
\bibitem [{\citenamefont {Li}\ \emph {et~al.}(2024{\natexlab{a}})\citenamefont {Li}, \citenamefont {Zhang}, \citenamefont {Xiang}, \citenamefont {Zhang}, \citenamefont {Zhu},\ and\ \citenamefont {Wen}}]{Li2024Signature}%
  \BibitemOpen
  \bibfield  {author} {\bibinfo {author} {\bibfnamefont {Q.}~\bibnamefont {Li}}, \bibinfo {author} {\bibfnamefont {Y.-J.}\ \bibnamefont {Zhang}}, \bibinfo {author} {\bibfnamefont {Z.-N.}\ \bibnamefont {Xiang}}, \bibinfo {author} {\bibfnamefont {Y.}~\bibnamefont {Zhang}}, \bibinfo {author} {\bibfnamefont {X.}~\bibnamefont {Zhu}}, \ and\ \bibinfo {author} {\bibfnamefont {H.-H.}\ \bibnamefont {Wen}},\ }\href {\doibase 10.1088/0256-307x/41/1/017401} {\bibfield  {journal} {\bibinfo  {journal} {Chinese Physics Letters}\ }\textbf {\bibinfo {volume} {41}},\ \bibinfo {pages} {017401} (\bibinfo {year} {2024}{\natexlab{a}})}\BibitemShut {NoStop}%
\bibitem [{\citenamefont {Zhang}\ \emph {et~al.}(2025{\natexlab{a}})\citenamefont {Zhang}, \citenamefont {Pei}, \citenamefont {Peng}, \citenamefont {Du}, \citenamefont {Hu}, \citenamefont {Cao}, \citenamefont {Wang}, \citenamefont {Wu}, \citenamefont {Li}, \citenamefont {Liu}, \citenamefont {Wen}, \citenamefont {Song}, \citenamefont {Zhao}, \citenamefont {Li}, \citenamefont {Cao}, \citenamefont {Zhu}, \citenamefont {Zhang}, \citenamefont {Yu}, \citenamefont {Cheng}, \citenamefont {Zhang}, \citenamefont {Li}, \citenamefont {Zhao}, \citenamefont {Chen}, \citenamefont {Jin}, \citenamefont {Guo}, \citenamefont {Wu}, \citenamefont {Yang}, \citenamefont {Zeng}, \citenamefont {Yan}, \citenamefont {Yang},\ and\ \citenamefont {Qi}}]{Zhang2024Superconductivity}%
  \BibitemOpen
  \bibfield  {author} {\bibinfo {author} {\bibfnamefont {M.}~\bibnamefont {Zhang}}, \bibinfo {author} {\bibfnamefont {C.}~\bibnamefont {Pei}}, \bibinfo {author} {\bibfnamefont {D.}~\bibnamefont {Peng}}, \bibinfo {author} {\bibfnamefont {X.}~\bibnamefont {Du}}, \bibinfo {author} {\bibfnamefont {W.}~\bibnamefont {Hu}}, \bibinfo {author} {\bibfnamefont {Y.}~\bibnamefont {Cao}}, \bibinfo {author} {\bibfnamefont {Q.}~\bibnamefont {Wang}}, \bibinfo {author} {\bibfnamefont {J.}~\bibnamefont {Wu}}, \bibinfo {author} {\bibfnamefont {Y.}~\bibnamefont {Li}}, \bibinfo {author} {\bibfnamefont {H.}~\bibnamefont {Liu}}, \bibinfo {author} {\bibfnamefont {C.}~\bibnamefont {Wen}}, \bibinfo {author} {\bibfnamefont {J.}~\bibnamefont {Song}}, \bibinfo {author} {\bibfnamefont {Y.}~\bibnamefont {Zhao}}, \bibinfo {author} {\bibfnamefont {C.}~\bibnamefont {Li}}, \bibinfo {author} {\bibfnamefont {W.}~\bibnamefont {Cao}}, \bibinfo {author} {\bibfnamefont {S.}~\bibnamefont {Zhu}}, \bibinfo {author} {\bibfnamefont {Q.}~\bibnamefont
  {Zhang}}, \bibinfo {author} {\bibfnamefont {N.}~\bibnamefont {Yu}}, \bibinfo {author} {\bibfnamefont {P.}~\bibnamefont {Cheng}}, \bibinfo {author} {\bibfnamefont {L.}~\bibnamefont {Zhang}}, \bibinfo {author} {\bibfnamefont {Z.}~\bibnamefont {Li}}, \bibinfo {author} {\bibfnamefont {J.}~\bibnamefont {Zhao}}, \bibinfo {author} {\bibfnamefont {Y.}~\bibnamefont {Chen}}, \bibinfo {author} {\bibfnamefont {C.}~\bibnamefont {Jin}}, \bibinfo {author} {\bibfnamefont {H.}~\bibnamefont {Guo}}, \bibinfo {author} {\bibfnamefont {C.}~\bibnamefont {Wu}}, \bibinfo {author} {\bibfnamefont {F.}~\bibnamefont {Yang}}, \bibinfo {author} {\bibfnamefont {Q.}~\bibnamefont {Zeng}}, \bibinfo {author} {\bibfnamefont {S.}~\bibnamefont {Yan}}, \bibinfo {author} {\bibfnamefont {L.}~\bibnamefont {Yang}}, \ and\ \bibinfo {author} {\bibfnamefont {Y.}~\bibnamefont {Qi}},\ }\href {\doibase 10.1103/PhysRevX.15.021005} {\bibfield  {journal} {\bibinfo  {journal} {Phys. Rev. X}\ }\textbf {\bibinfo {volume} {15}},\ \bibinfo {pages} {021005}
  (\bibinfo {year} {2025}{\natexlab{a}})}\BibitemShut {NoStop}%
\bibitem [{\citenamefont {Li}\ \emph {et~al.}(2024{\natexlab{b}})\citenamefont {Li}, \citenamefont {Chen}, \citenamefont {Huang}, \citenamefont {Han}, \citenamefont {Huo}, \citenamefont {Huang}, \citenamefont {Ma}, \citenamefont {Qiu}, \citenamefont {Chen}, \citenamefont {Hu}, \citenamefont {Chen}, \citenamefont {Xie}, \citenamefont {Shen}, \citenamefont {Sun}, \citenamefont {Yao},\ and\ \citenamefont {Wang}}]{Li2024Structural}%
  \BibitemOpen
  \bibfield  {author} {\bibinfo {author} {\bibfnamefont {J.}~\bibnamefont {Li}}, \bibinfo {author} {\bibfnamefont {C.-Q.}\ \bibnamefont {Chen}}, \bibinfo {author} {\bibfnamefont {C.}~\bibnamefont {Huang}}, \bibinfo {author} {\bibfnamefont {Y.}~\bibnamefont {Han}}, \bibinfo {author} {\bibfnamefont {M.}~\bibnamefont {Huo}}, \bibinfo {author} {\bibfnamefont {X.}~\bibnamefont {Huang}}, \bibinfo {author} {\bibfnamefont {P.}~\bibnamefont {Ma}}, \bibinfo {author} {\bibfnamefont {Z.}~\bibnamefont {Qiu}}, \bibinfo {author} {\bibfnamefont {J.}~\bibnamefont {Chen}}, \bibinfo {author} {\bibfnamefont {X.}~\bibnamefont {Hu}}, \bibinfo {author} {\bibfnamefont {L.}~\bibnamefont {Chen}}, \bibinfo {author} {\bibfnamefont {T.}~\bibnamefont {Xie}}, \bibinfo {author} {\bibfnamefont {B.}~\bibnamefont {Shen}}, \bibinfo {author} {\bibfnamefont {H.}~\bibnamefont {Sun}}, \bibinfo {author} {\bibfnamefont {D.-X.}\ \bibnamefont {Yao}}, \ and\ \bibinfo {author} {\bibfnamefont {M.}~\bibnamefont {Wang}},\ }\href
  {http://dx.doi.org/10.1007/s11433-023-2329-x} {\bibfield  {journal} {\bibinfo  {journal} {Science China Physics, Mechanics \& Astronomy}\ }\textbf {\bibinfo {volume} {67}},\ \bibinfo {pages} {117403} (\bibinfo {year} {2024}{\natexlab{b}})}\BibitemShut {NoStop}%
\bibitem [{\citenamefont {Carvalho}\ \emph {et~al.}(2000)\citenamefont {Carvalho}, \citenamefont {Cruz}, \citenamefont {Wattiaux}, \citenamefont {Bassat}, \citenamefont {Costa},\ and\ \citenamefont {Godinho}}]{Carvalho2000Influence}%
  \BibitemOpen
  \bibfield  {author} {\bibinfo {author} {\bibfnamefont {M.~D.}\ \bibnamefont {Carvalho}}, \bibinfo {author} {\bibfnamefont {M.~M.}\ \bibnamefont {Cruz}}, \bibinfo {author} {\bibfnamefont {A.}~\bibnamefont {Wattiaux}}, \bibinfo {author} {\bibfnamefont {J.~M.}\ \bibnamefont {Bassat}}, \bibinfo {author} {\bibfnamefont {F.~M.~A.}\ \bibnamefont {Costa}}, \ and\ \bibinfo {author} {\bibfnamefont {M.}~\bibnamefont {Godinho}},\ }\href {\doibase 10.1063/1.373693} {\bibfield  {journal} {\bibinfo  {journal} {Journal of Applied Physics}\ }\textbf {\bibinfo {volume} {88}},\ \bibinfo {pages} {544} (\bibinfo {year} {2000})}\BibitemShut {NoStop}%
\bibitem [{\citenamefont {Wu}\ \emph {et~al.}(2001)\citenamefont {Wu}, \citenamefont {Neumeier},\ and\ \citenamefont {Hundley}}]{Wu2001Magnetic}%
  \BibitemOpen
  \bibfield  {author} {\bibinfo {author} {\bibfnamefont {G.}~\bibnamefont {Wu}}, \bibinfo {author} {\bibfnamefont {J.~J.}\ \bibnamefont {Neumeier}}, \ and\ \bibinfo {author} {\bibfnamefont {M.~F.}\ \bibnamefont {Hundley}},\ }\href {\doibase 10.1103/PhysRevB.63.245120} {\bibfield  {journal} {\bibinfo  {journal} {Phys. Rev. B}\ }\textbf {\bibinfo {volume} {63}},\ \bibinfo {pages} {245120} (\bibinfo {year} {2001})}\BibitemShut {NoStop}%
\bibitem [{\citenamefont {Li}\ \emph {et~al.}(2017)\citenamefont {Li}, \citenamefont {Zhou}, \citenamefont {Nummy}, \citenamefont {Zhang}, \citenamefont {Pardo}, \citenamefont {Pickett}, \citenamefont {Mitchell},\ and\ \citenamefont {Dessau}}]{Li2017Fermiology}%
  \BibitemOpen
  \bibfield  {author} {\bibinfo {author} {\bibfnamefont {H.}~\bibnamefont {Li}}, \bibinfo {author} {\bibfnamefont {X.}~\bibnamefont {Zhou}}, \bibinfo {author} {\bibfnamefont {T.}~\bibnamefont {Nummy}}, \bibinfo {author} {\bibfnamefont {J.}~\bibnamefont {Zhang}}, \bibinfo {author} {\bibfnamefont {V.}~\bibnamefont {Pardo}}, \bibinfo {author} {\bibfnamefont {W.}~\bibnamefont {Pickett}}, \bibinfo {author} {\bibfnamefont {J.}~\bibnamefont {Mitchell}}, \ and\ \bibinfo {author} {\bibfnamefont {D.}~\bibnamefont {Dessau}},\ }\href {https://www.nature.com/articles/s41467-017-00777-0} {\bibfield  {journal} {\bibinfo  {journal} {Nature Communications}\ }\textbf {\bibinfo {volume} {8}},\ \bibinfo {pages} {704} (\bibinfo {year} {2017})}\BibitemShut {NoStop}%
\bibitem [{\citenamefont {Zhang}\ \emph {et~al.}(2020{\natexlab{a}})\citenamefont {Zhang}, \citenamefont {Zheng}, \citenamefont {Chen}, \citenamefont {Ren}, \citenamefont {Yonemura}, \citenamefont {Huq},\ and\ \citenamefont {Mitchell}}]{Zhang2020High}%
  \BibitemOpen
  \bibfield  {author} {\bibinfo {author} {\bibfnamefont {J.}~\bibnamefont {Zhang}}, \bibinfo {author} {\bibfnamefont {H.}~\bibnamefont {Zheng}}, \bibinfo {author} {\bibfnamefont {Y.-S.}\ \bibnamefont {Chen}}, \bibinfo {author} {\bibfnamefont {Y.}~\bibnamefont {Ren}}, \bibinfo {author} {\bibfnamefont {M.}~\bibnamefont {Yonemura}}, \bibinfo {author} {\bibfnamefont {A.}~\bibnamefont {Huq}}, \ and\ \bibinfo {author} {\bibfnamefont {J.~F.}\ \bibnamefont {Mitchell}},\ }\href {\doibase 10.1103/PhysRevMaterials.4.083402} {\bibfield  {journal} {\bibinfo  {journal} {Phys. Rev. Mater.}\ }\textbf {\bibinfo {volume} {4}},\ \bibinfo {pages} {083402} (\bibinfo {year} {2020}{\natexlab{a}})}\BibitemShut {NoStop}%
\bibitem [{\citenamefont {Kumar}\ \emph {et~al.}(2020)\citenamefont {Kumar}, \citenamefont {Fjellv{\aa}g}, \citenamefont {Sj{\aa}stad},\ and\ \citenamefont {Fjellv{\aa}g}}]{Susmit2020Physical}%
  \BibitemOpen
  \bibfield  {author} {\bibinfo {author} {\bibfnamefont {S.}~\bibnamefont {Kumar}}, \bibinfo {author} {\bibfnamefont {{\O}.}~\bibnamefont {Fjellv{\aa}g}}, \bibinfo {author} {\bibfnamefont {A.~O.}\ \bibnamefont {Sj{\aa}stad}}, \ and\ \bibinfo {author} {\bibfnamefont {H.}~\bibnamefont {Fjellv{\aa}g}},\ }\href {\doibase https://doi.org/10.1016/j.jmmm.2019.165915} {\bibfield  {journal} {\bibinfo  {journal} {Journal of Magnetism and Magnetic Materials}\ }\textbf {\bibinfo {volume} {496}},\ \bibinfo {pages} {165915} (\bibinfo {year} {2020})}\BibitemShut {NoStop}%
\bibitem [{\citenamefont {Li}\ \emph {et~al.}(2020)\citenamefont {Li}, \citenamefont {Wang}, \citenamefont {Yang}, \citenamefont {Sun}, \citenamefont {Liu}, \citenamefont {Wu}, \citenamefont {Ren}, \citenamefont {Cheng}, \citenamefont {Zhang},\ and\ \citenamefont {Cao}}]{Li2020Metal-to-metal}%
  \BibitemOpen
  \bibfield  {author} {\bibinfo {author} {\bibfnamefont {B.-Z.}\ \bibnamefont {Li}}, \bibinfo {author} {\bibfnamefont {C.}~\bibnamefont {Wang}}, \bibinfo {author} {\bibfnamefont {P.~T.}\ \bibnamefont {Yang}}, \bibinfo {author} {\bibfnamefont {J.~P.}\ \bibnamefont {Sun}}, \bibinfo {author} {\bibfnamefont {Y.-B.}\ \bibnamefont {Liu}}, \bibinfo {author} {\bibfnamefont {J.}~\bibnamefont {Wu}}, \bibinfo {author} {\bibfnamefont {Z.}~\bibnamefont {Ren}}, \bibinfo {author} {\bibfnamefont {J.-G.}\ \bibnamefont {Cheng}}, \bibinfo {author} {\bibfnamefont {G.-M.}\ \bibnamefont {Zhang}}, \ and\ \bibinfo {author} {\bibfnamefont {G.-H.}\ \bibnamefont {Cao}},\ }\href {\doibase 10.1103/PhysRevB.101.195142} {\bibfield  {journal} {\bibinfo  {journal} {Phys. Rev. B}\ }\textbf {\bibinfo {volume} {101}},\ \bibinfo {pages} {195142} (\bibinfo {year} {2020})}\BibitemShut {NoStop}%
\bibitem [{\citenamefont {Huangfu}\ \emph {et~al.}(2020)\citenamefont {Huangfu}, \citenamefont {Zhang},\ and\ \citenamefont {Schilling}}]{Huang2020Correlation}%
  \BibitemOpen
  \bibfield  {author} {\bibinfo {author} {\bibfnamefont {S.}~\bibnamefont {Huangfu}}, \bibinfo {author} {\bibfnamefont {X.}~\bibnamefont {Zhang}}, \ and\ \bibinfo {author} {\bibfnamefont {A.}~\bibnamefont {Schilling}},\ }\href {\doibase 10.1103/PhysRevResearch.2.033247} {\bibfield  {journal} {\bibinfo  {journal} {Phys. Rev. Res.}\ }\textbf {\bibinfo {volume} {2}},\ \bibinfo {pages} {033247} (\bibinfo {year} {2020})}\BibitemShut {NoStop}%
\bibitem [{\citenamefont {Rout}\ \emph {et~al.}(2020)\citenamefont {Rout}, \citenamefont {Mudi}, \citenamefont {Hoffmann}, \citenamefont {Spachmann}, \citenamefont {Klingeler},\ and\ \citenamefont {Singh}}]{Rout2020Structural}%
  \BibitemOpen
  \bibfield  {author} {\bibinfo {author} {\bibfnamefont {D.}~\bibnamefont {Rout}}, \bibinfo {author} {\bibfnamefont {S.~R.}\ \bibnamefont {Mudi}}, \bibinfo {author} {\bibfnamefont {M.}~\bibnamefont {Hoffmann}}, \bibinfo {author} {\bibfnamefont {S.}~\bibnamefont {Spachmann}}, \bibinfo {author} {\bibfnamefont {R.}~\bibnamefont {Klingeler}}, \ and\ \bibinfo {author} {\bibfnamefont {S.}~\bibnamefont {Singh}},\ }\href {\doibase 10.1103/PhysRevB.102.195144} {\bibfield  {journal} {\bibinfo  {journal} {Phys. Rev. B}\ }\textbf {\bibinfo {volume} {102}},\ \bibinfo {pages} {195144} (\bibinfo {year} {2020})}\BibitemShut {NoStop}%
\bibitem [{\citenamefont {Zhang}\ \emph {et~al.}(2020{\natexlab{b}})\citenamefont {Zhang}, \citenamefont {Phelan}, \citenamefont {Botana}, \citenamefont {Chen}, \citenamefont {Zheng}, \citenamefont {Krogstad}, \citenamefont {Wang}, \citenamefont {Qiu}, \citenamefont {Rodriguez-Rivera}, \citenamefont {Osborn}, \citenamefont {Rosenkranz}, \citenamefont {Norman},\ and\ \citenamefont {Mitchell}}]{Zhang2020Interwined}%
  \BibitemOpen
  \bibfield  {author} {\bibinfo {author} {\bibfnamefont {J.}~\bibnamefont {Zhang}}, \bibinfo {author} {\bibfnamefont {D.}~\bibnamefont {Phelan}}, \bibinfo {author} {\bibfnamefont {A.~S.}\ \bibnamefont {Botana}}, \bibinfo {author} {\bibfnamefont {Y.-S.}\ \bibnamefont {Chen}}, \bibinfo {author} {\bibfnamefont {H.}~\bibnamefont {Zheng}}, \bibinfo {author} {\bibfnamefont {M.}~\bibnamefont {Krogstad}}, \bibinfo {author} {\bibfnamefont {S.~G.}\ \bibnamefont {Wang}}, \bibinfo {author} {\bibfnamefont {Y.}~\bibnamefont {Qiu}}, \bibinfo {author} {\bibfnamefont {J.~A.}\ \bibnamefont {Rodriguez-Rivera}}, \bibinfo {author} {\bibfnamefont {R.}~\bibnamefont {Osborn}}, \bibinfo {author} {\bibfnamefont {S.}~\bibnamefont {Rosenkranz}}, \bibinfo {author} {\bibfnamefont {M.~R.}\ \bibnamefont {Norman}}, \ and\ \bibinfo {author} {\bibfnamefont {J.~F.}\ \bibnamefont {Mitchell}},\ }\href {http://dx.doi.org/10.1038/s41467-020-19836-0} {\bibfield  {journal} {\bibinfo  {journal} {Nature Communications}\ }\textbf {\bibinfo {volume}
  {11}},\ \bibinfo {pages} {6003} (\bibinfo {year} {2020}{\natexlab{b}})}\BibitemShut {NoStop}%
\bibitem [{\citenamefont {Yuan}\ \emph {et~al.}(2024)\citenamefont {Yuan}, \citenamefont {Elghandour}, \citenamefont {Arneth}, \citenamefont {Dey},\ and\ \citenamefont {Klingeler}}]{Ning2024High-pressure}%
  \BibitemOpen
  \bibfield  {author} {\bibinfo {author} {\bibfnamefont {N.}~\bibnamefont {Yuan}}, \bibinfo {author} {\bibfnamefont {A.}~\bibnamefont {Elghandour}}, \bibinfo {author} {\bibfnamefont {J.}~\bibnamefont {Arneth}}, \bibinfo {author} {\bibfnamefont {K.}~\bibnamefont {Dey}}, \ and\ \bibinfo {author} {\bibfnamefont {R.}~\bibnamefont {Klingeler}},\ }\href {\doibase https://doi.org/10.1016/j.jcrysgro.2023.127511} {\bibfield  {journal} {\bibinfo  {journal} {Journal of Crystal Growth}\ }\textbf {\bibinfo {volume} {627}},\ \bibinfo {pages} {127511} (\bibinfo {year} {2024})}\BibitemShut {NoStop}%
\bibitem [{\citenamefont {Kakoi}\ \emph {et~al.}(2024)\citenamefont {Kakoi}, \citenamefont {Oi}, \citenamefont {Ohshita}, \citenamefont {Yashima}, \citenamefont {Kuroki}, \citenamefont {Kato}, \citenamefont {Takahashi}, \citenamefont {Ishiwata}, \citenamefont {Adachi}, \citenamefont {Hatada}, \citenamefont {Uda},\ and\ \citenamefont {Mukuda}}]{Kakoi2024Multiband}%
  \BibitemOpen
  \bibfield  {author} {\bibinfo {author} {\bibfnamefont {M.}~\bibnamefont {Kakoi}}, \bibinfo {author} {\bibfnamefont {T.}~\bibnamefont {Oi}}, \bibinfo {author} {\bibfnamefont {Y.}~\bibnamefont {Ohshita}}, \bibinfo {author} {\bibfnamefont {M.}~\bibnamefont {Yashima}}, \bibinfo {author} {\bibfnamefont {K.}~\bibnamefont {Kuroki}}, \bibinfo {author} {\bibfnamefont {T.}~\bibnamefont {Kato}}, \bibinfo {author} {\bibfnamefont {H.}~\bibnamefont {Takahashi}}, \bibinfo {author} {\bibfnamefont {S.}~\bibnamefont {Ishiwata}}, \bibinfo {author} {\bibfnamefont {Y.}~\bibnamefont {Adachi}}, \bibinfo {author} {\bibfnamefont {N.}~\bibnamefont {Hatada}}, \bibinfo {author} {\bibfnamefont {T.}~\bibnamefont {Uda}}, \ and\ \bibinfo {author} {\bibfnamefont {H.}~\bibnamefont {Mukuda}},\ }\href {https://doi.org/10.7566/JPSJ.93.053702} {\bibfield  {journal} {\bibinfo  {journal} {Journal of the Physical Society of Japan}\ }\textbf {\bibinfo {volume} {93}},\ \bibinfo {pages} {053702} (\bibinfo {year} {2024})}\BibitemShut {NoStop}%
\bibitem [{\citenamefont {Leonov}(2024)}]{Leonov2024Electronic}%
  \BibitemOpen
  \bibfield  {author} {\bibinfo {author} {\bibfnamefont {I.~V.}\ \bibnamefont {Leonov}},\ }\href {\doibase 10.1103/PhysRevB.109.235123} {\bibfield  {journal} {\bibinfo  {journal} {Phys. Rev. B}\ }\textbf {\bibinfo {volume} {109}},\ \bibinfo {pages} {235123} (\bibinfo {year} {2024})}\BibitemShut {NoStop}%
\bibitem [{\citenamefont {Wang}\ \emph {et~al.}(2024{\natexlab{a}})\citenamefont {Wang}, \citenamefont {Ouyang}, \citenamefont {He},\ and\ \citenamefont {Lu}}]{Wang2024Non-Fermi}%
  \BibitemOpen
  \bibfield  {author} {\bibinfo {author} {\bibfnamefont {J.-X.}\ \bibnamefont {Wang}}, \bibinfo {author} {\bibfnamefont {Z.}~\bibnamefont {Ouyang}}, \bibinfo {author} {\bibfnamefont {R.-Q.}\ \bibnamefont {He}}, \ and\ \bibinfo {author} {\bibfnamefont {Z.-Y.}\ \bibnamefont {Lu}},\ }\href {\doibase 10.1103/PhysRevB.109.165140} {\bibfield  {journal} {\bibinfo  {journal} {Phys. Rev. B}\ }\textbf {\bibinfo {volume} {109}},\ \bibinfo {pages} {165140} (\bibinfo {year} {2024}{\natexlab{a}})}\BibitemShut {NoStop}%
\bibitem [{\citenamefont {Tian}\ \emph {et~al.}(2024)\citenamefont {Tian}, \citenamefont {Ma}, \citenamefont {Ming}, \citenamefont {Zheng},\ and\ \citenamefont {Li}}]{Tian2024Effective}%
  \BibitemOpen
  \bibfield  {author} {\bibinfo {author} {\bibfnamefont {P.-F.}\ \bibnamefont {Tian}}, \bibinfo {author} {\bibfnamefont {H.-T.}\ \bibnamefont {Ma}}, \bibinfo {author} {\bibfnamefont {X.}~\bibnamefont {Ming}}, \bibinfo {author} {\bibfnamefont {X.-J.}\ \bibnamefont {Zheng}}, \ and\ \bibinfo {author} {\bibfnamefont {H.}~\bibnamefont {Li}},\ }\href {\doibase 10.1088/1361-648X/ad512c} {\bibfield  {journal} {\bibinfo  {journal} {Journal of Physics: Condensed Matter}\ }\textbf {\bibinfo {volume} {36}},\ \bibinfo {pages} {355602} (\bibinfo {year} {2024})}\BibitemShut {NoStop}%
\bibitem [{\citenamefont {LaBollita}\ \emph {et~al.}(2024)\citenamefont {LaBollita}, \citenamefont {Kapeghian}, \citenamefont {Norman},\ and\ \citenamefont {Botana}}]{LaBollita2024Electronic}%
  \BibitemOpen
  \bibfield  {author} {\bibinfo {author} {\bibfnamefont {H.}~\bibnamefont {LaBollita}}, \bibinfo {author} {\bibfnamefont {J.}~\bibnamefont {Kapeghian}}, \bibinfo {author} {\bibfnamefont {M.~R.}\ \bibnamefont {Norman}}, \ and\ \bibinfo {author} {\bibfnamefont {A.~S.}\ \bibnamefont {Botana}},\ }\href {\doibase 10.1103/PhysRevB.109.195151} {\bibfield  {journal} {\bibinfo  {journal} {Phys. Rev. B}\ }\textbf {\bibinfo {volume} {109}},\ \bibinfo {pages} {195151} (\bibinfo {year} {2024})}\BibitemShut {NoStop}%
\bibitem [{\citenamefont {Yang}\ \emph {et~al.}(2024)\citenamefont {Yang}, \citenamefont {Jiang}, \citenamefont {Wang}, \citenamefont {Lu},\ and\ \citenamefont {Wang}}]{Yang2024Effective}%
  \BibitemOpen
  \bibfield  {author} {\bibinfo {author} {\bibfnamefont {Q.-G.}\ \bibnamefont {Yang}}, \bibinfo {author} {\bibfnamefont {K.-Y.}\ \bibnamefont {Jiang}}, \bibinfo {author} {\bibfnamefont {D.}~\bibnamefont {Wang}}, \bibinfo {author} {\bibfnamefont {H.-Y.}\ \bibnamefont {Lu}}, \ and\ \bibinfo {author} {\bibfnamefont {Q.-H.}\ \bibnamefont {Wang}},\ }\href {\doibase 10.1103/PhysRevB.109.L220506} {\bibfield  {journal} {\bibinfo  {journal} {Phys. Rev. B}\ }\textbf {\bibinfo {volume} {109}},\ \bibinfo {pages} {L220506} (\bibinfo {year} {2024})}\BibitemShut {NoStop}%
\bibitem [{\citenamefont {Chen}\ \emph {et~al.}(2024)\citenamefont {Chen}, \citenamefont {Luo}, \citenamefont {Wang}, \citenamefont {W\'u},\ and\ \citenamefont {Yao}}]{Chen2024Trilayer}%
  \BibitemOpen
  \bibfield  {author} {\bibinfo {author} {\bibfnamefont {C.-Q.}\ \bibnamefont {Chen}}, \bibinfo {author} {\bibfnamefont {Z.}~\bibnamefont {Luo}}, \bibinfo {author} {\bibfnamefont {M.}~\bibnamefont {Wang}}, \bibinfo {author} {\bibfnamefont {W.}~\bibnamefont {W\'u}}, \ and\ \bibinfo {author} {\bibfnamefont {D.-X.}\ \bibnamefont {Yao}},\ }\href {\doibase 10.1103/PhysRevB.110.014503} {\bibfield  {journal} {\bibinfo  {journal} {Phys. Rev. B}\ }\textbf {\bibinfo {volume} {110}},\ \bibinfo {pages} {014503} (\bibinfo {year} {2024})}\BibitemShut {NoStop}%
\bibitem [{\citenamefont {Du}\ \emph {et~al.}(2024)\citenamefont {Du}, \citenamefont {Li}, \citenamefont {Cao}, \citenamefont {Pei}, \citenamefont {Zhang}, \citenamefont {Zhao}, \citenamefont {Zhai}, \citenamefont {Xu}, \citenamefont {Liu}, \citenamefont {Li}, \citenamefont {Zhao}, \citenamefont {Li}, \citenamefont {Chen}, \citenamefont {Qi}, \citenamefont {Guo},\ and\ \citenamefont {Yang}}]{Du2024correlated}%
  \BibitemOpen
  \bibfield  {author} {\bibinfo {author} {\bibfnamefont {X.}~\bibnamefont {Du}}, \bibinfo {author} {\bibfnamefont {Y.~D.}\ \bibnamefont {Li}}, \bibinfo {author} {\bibfnamefont {Y.~T.}\ \bibnamefont {Cao}}, \bibinfo {author} {\bibfnamefont {C.~Y.}\ \bibnamefont {Pei}}, \bibinfo {author} {\bibfnamefont {M.~X.}\ \bibnamefont {Zhang}}, \bibinfo {author} {\bibfnamefont {W.~X.}\ \bibnamefont {Zhao}}, \bibinfo {author} {\bibfnamefont {K.~Y.}\ \bibnamefont {Zhai}}, \bibinfo {author} {\bibfnamefont {R.~Z.}\ \bibnamefont {Xu}}, \bibinfo {author} {\bibfnamefont {Z.~K.}\ \bibnamefont {Liu}}, \bibinfo {author} {\bibfnamefont {Z.~W.}\ \bibnamefont {Li}}, \bibinfo {author} {\bibfnamefont {J.~K.}\ \bibnamefont {Zhao}}, \bibinfo {author} {\bibfnamefont {G.}~\bibnamefont {Li}}, \bibinfo {author} {\bibfnamefont {Y.~L.}\ \bibnamefont {Chen}}, \bibinfo {author} {\bibfnamefont {Y.~P.}\ \bibnamefont {Qi}}, \bibinfo {author} {\bibfnamefont {H.~J.}\ \bibnamefont {Guo}}, \ and\ \bibinfo {author} {\bibfnamefont {L.~X.}\ \bibnamefont
  {Yang}},\ }\href {https://arxiv.org/abs/2405.19853} {\bibfield  {journal} {\bibinfo  {journal} {arXiv:2405.19853}\ } (\bibinfo {year} {2024})}\BibitemShut {NoStop}%
\bibitem [{\citenamefont {Xu}\ \emph {et~al.}(2025{\natexlab{a}})\citenamefont {Xu}, \citenamefont {Chen}, \citenamefont {Huo}, \citenamefont {Hu}, \citenamefont {Wang}, \citenamefont {Wu}, \citenamefont {Li}, \citenamefont {Wu}, \citenamefont {Wang}, \citenamefont {Yao}, \citenamefont {Dong},\ and\ \citenamefont {Wang}}]{Xu2024Origin}%
  \BibitemOpen
  \bibfield  {author} {\bibinfo {author} {\bibfnamefont {S.}~\bibnamefont {Xu}}, \bibinfo {author} {\bibfnamefont {C.-Q.}\ \bibnamefont {Chen}}, \bibinfo {author} {\bibfnamefont {M.}~\bibnamefont {Huo}}, \bibinfo {author} {\bibfnamefont {D.}~\bibnamefont {Hu}}, \bibinfo {author} {\bibfnamefont {H.}~\bibnamefont {Wang}}, \bibinfo {author} {\bibfnamefont {Q.}~\bibnamefont {Wu}}, \bibinfo {author} {\bibfnamefont {R.}~\bibnamefont {Li}}, \bibinfo {author} {\bibfnamefont {D.}~\bibnamefont {Wu}}, \bibinfo {author} {\bibfnamefont {M.}~\bibnamefont {Wang}}, \bibinfo {author} {\bibfnamefont {D.-X.}\ \bibnamefont {Yao}}, \bibinfo {author} {\bibfnamefont {T.}~\bibnamefont {Dong}}, \ and\ \bibinfo {author} {\bibfnamefont {N.}~\bibnamefont {Wang}},\ }\href {\doibase 10.1103/PhysRevB.111.075140} {\bibfield  {journal} {\bibinfo  {journal} {Phys. Rev. B}\ }\textbf {\bibinfo {volume} {111}},\ \bibinfo {pages} {075140} (\bibinfo {year} {2025}{\natexlab{a}})}\BibitemShut {NoStop}%
\bibitem [{\citenamefont {Qin}\ \emph {et~al.}(2024)\citenamefont {Qin}, \citenamefont {Wang},\ and\ \citenamefont {Yang}}]{Qin2024Frustrated}%
  \BibitemOpen
  \bibfield  {author} {\bibinfo {author} {\bibfnamefont {Q.}~\bibnamefont {Qin}}, \bibinfo {author} {\bibfnamefont {J.}~\bibnamefont {Wang}}, \ and\ \bibinfo {author} {\bibfnamefont {Y.-f.}\ \bibnamefont {Yang}},\ }\href {\doibase 10.59717/j.xinn-mater.2024.100102} {\bibfield  {journal} {\bibinfo  {journal} {The Innovation Materials}\ }\textbf {\bibinfo {volume} {2}},\ \bibinfo {pages} {100102} (\bibinfo {year} {2024})}\BibitemShut {NoStop}%
\bibitem [{\citenamefont {Li}\ \emph {et~al.}(2025{\natexlab{a}})\citenamefont {Li}, \citenamefont {Cao}, \citenamefont {Liu}, \citenamefont {Peng}, \citenamefont {Lin}, \citenamefont {Pei}, \citenamefont {Zhang}, \citenamefont {Wu}, \citenamefont {Du}, \citenamefont {Zhao}, \citenamefont {Zhai}, \citenamefont {Zhang}, \citenamefont {Zhao}, \citenamefont {Lin}, \citenamefont {Tan}, \citenamefont {Qi}, \citenamefont {Li}, \citenamefont {Guo}, \citenamefont {Yang},\ and\ \citenamefont {Yang}}]{Li2025Distinct}%
  \BibitemOpen
  \bibfield  {author} {\bibinfo {author} {\bibfnamefont {Y.}~\bibnamefont {Li}}, \bibinfo {author} {\bibfnamefont {Y.}~\bibnamefont {Cao}}, \bibinfo {author} {\bibfnamefont {L.}~\bibnamefont {Liu}}, \bibinfo {author} {\bibfnamefont {P.}~\bibnamefont {Peng}}, \bibinfo {author} {\bibfnamefont {H.}~\bibnamefont {Lin}}, \bibinfo {author} {\bibfnamefont {C.}~\bibnamefont {Pei}}, \bibinfo {author} {\bibfnamefont {M.}~\bibnamefont {Zhang}}, \bibinfo {author} {\bibfnamefont {H.}~\bibnamefont {Wu}}, \bibinfo {author} {\bibfnamefont {X.}~\bibnamefont {Du}}, \bibinfo {author} {\bibfnamefont {W.}~\bibnamefont {Zhao}}, \bibinfo {author} {\bibfnamefont {K.}~\bibnamefont {Zhai}}, \bibinfo {author} {\bibfnamefont {X.}~\bibnamefont {Zhang}}, \bibinfo {author} {\bibfnamefont {J.}~\bibnamefont {Zhao}}, \bibinfo {author} {\bibfnamefont {M.}~\bibnamefont {Lin}}, \bibinfo {author} {\bibfnamefont {P.}~\bibnamefont {Tan}}, \bibinfo {author} {\bibfnamefont {Y.}~\bibnamefont {Qi}}, \bibinfo {author} {\bibfnamefont {G.}~\bibnamefont
  {Li}}, \bibinfo {author} {\bibfnamefont {H.}~\bibnamefont {Guo}}, \bibinfo {author} {\bibfnamefont {L.}~\bibnamefont {Yang}}, \ and\ \bibinfo {author} {\bibfnamefont {L.}~\bibnamefont {Yang}},\ }\href {\doibase https://doi.org/10.1016/j.scib.2024.10.011} {\bibfield  {journal} {\bibinfo  {journal} {Science Bulletin}\ }\textbf {\bibinfo {volume} {70}},\ \bibinfo {pages} {180} (\bibinfo {year} {2025}{\natexlab{a}})}\BibitemShut {NoStop}%
\bibitem [{\citenamefont {Zhang}\ \emph {et~al.}(2024{\natexlab{b}})\citenamefont {Zhang}, \citenamefont {Lin}, \citenamefont {Moreo}, \citenamefont {Maier},\ and\ \citenamefont {Dagotto}}]{Zhang2024Prediction}%
  \BibitemOpen
  \bibfield  {author} {\bibinfo {author} {\bibfnamefont {Y.}~\bibnamefont {Zhang}}, \bibinfo {author} {\bibfnamefont {L.-F.}\ \bibnamefont {Lin}}, \bibinfo {author} {\bibfnamefont {A.}~\bibnamefont {Moreo}}, \bibinfo {author} {\bibfnamefont {T.~A.}\ \bibnamefont {Maier}}, \ and\ \bibinfo {author} {\bibfnamefont {E.}~\bibnamefont {Dagotto}},\ }\href {\doibase 10.1103/PhysRevLett.133.136001} {\bibfield  {journal} {\bibinfo  {journal} {Phys. Rev. Lett.}\ }\textbf {\bibinfo {volume} {133}},\ \bibinfo {pages} {136001} (\bibinfo {year} {2024}{\natexlab{b}})}\BibitemShut {NoStop}%
\bibitem [{\citenamefont {Deswal}\ \emph {et~al.}(2024)\citenamefont {Deswal}, \citenamefont {Kumar}, \citenamefont {Rout}, \citenamefont {Singh},\ and\ \citenamefont {Kumar}}]{Sonia2024Dynamics}%
  \BibitemOpen
  \bibfield  {author} {\bibinfo {author} {\bibfnamefont {S.}~\bibnamefont {Deswal}}, \bibinfo {author} {\bibfnamefont {D.}~\bibnamefont {Kumar}}, \bibinfo {author} {\bibfnamefont {D.}~\bibnamefont {Rout}}, \bibinfo {author} {\bibfnamefont {S.}~\bibnamefont {Singh}}, \ and\ \bibinfo {author} {\bibfnamefont {P.}~\bibnamefont {Kumar}},\ }\href {https://arxiv.org/abs/2411.13933} {\bibfield  {journal} {\bibinfo  {journal} {arXiv:2411.13933}\ } (\bibinfo {year} {2024})}\BibitemShut {NoStop}%
\bibitem [{\citenamefont {Zhang}\ \emph {et~al.}(2025{\natexlab{b}})\citenamefont {Zhang}, \citenamefont {Peng}, \citenamefont {Zhu}, \citenamefont {Chen}, \citenamefont {Cui}, \citenamefont {Wang}, \citenamefont {Wang}, \citenamefont {Zeng},\ and\ \citenamefont {Zhao}}]{Zhang2025Bulk}%
  \BibitemOpen
  \bibfield  {author} {\bibinfo {author} {\bibfnamefont {E.}~\bibnamefont {Zhang}}, \bibinfo {author} {\bibfnamefont {D.}~\bibnamefont {Peng}}, \bibinfo {author} {\bibfnamefont {Y.}~\bibnamefont {Zhu}}, \bibinfo {author} {\bibfnamefont {L.}~\bibnamefont {Chen}}, \bibinfo {author} {\bibfnamefont {B.}~\bibnamefont {Cui}}, \bibinfo {author} {\bibfnamefont {X.}~\bibnamefont {Wang}}, \bibinfo {author} {\bibfnamefont {W.}~\bibnamefont {Wang}}, \bibinfo {author} {\bibfnamefont {Q.}~\bibnamefont {Zeng}}, \ and\ \bibinfo {author} {\bibfnamefont {J.}~\bibnamefont {Zhao}},\ }\href {\doibase 10.1103/PhysRevX.15.021008} {\bibfield  {journal} {\bibinfo  {journal} {Phys. Rev. X}\ }\textbf {\bibinfo {volume} {15}},\ \bibinfo {pages} {021008} (\bibinfo {year} {2025}{\natexlab{b}})}\BibitemShut {NoStop}%
\bibitem [{\citenamefont {Li}\ \emph {et~al.}(2025{\natexlab{b}})\citenamefont {Li}, \citenamefont {Gong}, \citenamefont {Zhu}, \citenamefont {Chen}, \citenamefont {Zhang}, \citenamefont {Zhang}, \citenamefont {Li}, \citenamefont {Yin}, \citenamefont {Wang}, \citenamefont {Zhao}, \citenamefont {Feng}, \citenamefont {Du},\ and\ \citenamefont {Yan}}]{Mingzhe2025Direct}%
  \BibitemOpen
  \bibfield  {author} {\bibinfo {author} {\bibfnamefont {M.}~\bibnamefont {Li}}, \bibinfo {author} {\bibfnamefont {J.}~\bibnamefont {Gong}}, \bibinfo {author} {\bibfnamefont {Y.}~\bibnamefont {Zhu}}, \bibinfo {author} {\bibfnamefont {Z.}~\bibnamefont {Chen}}, \bibinfo {author} {\bibfnamefont {J.}~\bibnamefont {Zhang}}, \bibinfo {author} {\bibfnamefont {E.}~\bibnamefont {Zhang}}, \bibinfo {author} {\bibfnamefont {Y.}~\bibnamefont {Li}}, \bibinfo {author} {\bibfnamefont {R.}~\bibnamefont {Yin}}, \bibinfo {author} {\bibfnamefont {S.}~\bibnamefont {Wang}}, \bibinfo {author} {\bibfnamefont {J.}~\bibnamefont {Zhao}}, \bibinfo {author} {\bibfnamefont {D.-L.}\ \bibnamefont {Feng}}, \bibinfo {author} {\bibfnamefont {Z.}~\bibnamefont {Du}}, \ and\ \bibinfo {author} {\bibfnamefont {Y.-J.}\ \bibnamefont {Yan}},\ }\href {https://arxiv.org/abs/2501.18885} {\bibfield  {journal} {\bibinfo  {journal} {arXiv:2501.18885}\ } (\bibinfo {year} {2025}{\natexlab{b}})}\BibitemShut {NoStop}%
\bibitem [{\citenamefont {Zhang}(2025)}]{Hu2025Origin}%
  \BibitemOpen
  \bibfield  {author} {\bibinfo {author} {\bibfnamefont {H.}~\bibnamefont {Zhang}},\ }\href {https://arxiv.org/abs/2502.13354} {\bibfield  {journal} {\bibinfo  {journal} {arXiv:2502.13354}\ } (\bibinfo {year} {2025})}\BibitemShut {NoStop}%
\bibitem [{\citenamefont {Peng}\ \emph {et~al.}(2025)\citenamefont {Peng}, \citenamefont {Bian}, \citenamefont {Xing}, \citenamefont {Chen}, \citenamefont {Cai}, \citenamefont {Luo}, \citenamefont {Lan}, \citenamefont {Liu}, \citenamefont {Zhu}, \citenamefont {Zhang}, \citenamefont {Wang}, \citenamefont {Sun}, \citenamefont {Wang}, \citenamefont {Wang}, \citenamefont {Wang}, \citenamefont {Yang}, \citenamefont {Yang}, \citenamefont {Dong}, \citenamefont {Lou}, \citenamefont {Zeng}, \citenamefont {Zeng}, \citenamefont {Tian}, \citenamefont {Zhao}, \citenamefont {Zeng}, \citenamefont {Zhang},\ and\ \citenamefont {kwang Mao}}]{Di2025Isotropic}%
  \BibitemOpen
  \bibfield  {author} {\bibinfo {author} {\bibfnamefont {D.}~\bibnamefont {Peng}}, \bibinfo {author} {\bibfnamefont {Y.}~\bibnamefont {Bian}}, \bibinfo {author} {\bibfnamefont {Z.}~\bibnamefont {Xing}}, \bibinfo {author} {\bibfnamefont {L.}~\bibnamefont {Chen}}, \bibinfo {author} {\bibfnamefont {J.}~\bibnamefont {Cai}}, \bibinfo {author} {\bibfnamefont {T.}~\bibnamefont {Luo}}, \bibinfo {author} {\bibfnamefont {F.}~\bibnamefont {Lan}}, \bibinfo {author} {\bibfnamefont {Y.}~\bibnamefont {Liu}}, \bibinfo {author} {\bibfnamefont {Y.}~\bibnamefont {Zhu}}, \bibinfo {author} {\bibfnamefont {E.}~\bibnamefont {Zhang}}, \bibinfo {author} {\bibfnamefont {Z.}~\bibnamefont {Wang}}, \bibinfo {author} {\bibfnamefont {Y.}~\bibnamefont {Sun}}, \bibinfo {author} {\bibfnamefont {Y.}~\bibnamefont {Wang}}, \bibinfo {author} {\bibfnamefont {X.}~\bibnamefont {Wang}}, \bibinfo {author} {\bibfnamefont {C.}~\bibnamefont {Wang}}, \bibinfo {author} {\bibfnamefont {Y.}~\bibnamefont {Yang}}, \bibinfo {author} {\bibfnamefont
  {Y.}~\bibnamefont {Yang}}, \bibinfo {author} {\bibfnamefont {H.}~\bibnamefont {Dong}}, \bibinfo {author} {\bibfnamefont {H.}~\bibnamefont {Lou}}, \bibinfo {author} {\bibfnamefont {Z.}~\bibnamefont {Zeng}}, \bibinfo {author} {\bibfnamefont {Z.}~\bibnamefont {Zeng}}, \bibinfo {author} {\bibfnamefont {M.}~\bibnamefont {Tian}}, \bibinfo {author} {\bibfnamefont {J.}~\bibnamefont {Zhao}}, \bibinfo {author} {\bibfnamefont {Q.}~\bibnamefont {Zeng}}, \bibinfo {author} {\bibfnamefont {J.}~\bibnamefont {Zhang}}, \ and\ \bibinfo {author} {\bibfnamefont {H.}~\bibnamefont {kwang Mao}},\ }\href {https://arxiv.org/abs/2502.14410} {\bibfield  {journal} {\bibinfo  {journal} {arXiv:2502.14410}\ } (\bibinfo {year} {2025})}\BibitemShut {NoStop}%
\bibitem [{\citenamefont {Khasanov}\ \emph {et~al.}(2025)\citenamefont {Khasanov}, \citenamefont {Hicken}, \citenamefont {Plokhikh}, \citenamefont {Sazgari}, \citenamefont {Keller}, \citenamefont {Pomjakushin}, \citenamefont {Bartkowiak}, \citenamefont {Kr\'{o}lak}, \citenamefont {Winiarski}, \citenamefont {Krieger}, \citenamefont {Luetkens}, \citenamefont {Klimczuk}, \citenamefont {Gawryluk},\ and\ \citenamefont {Guguchia}}]{khasanov2025identical}%
  \BibitemOpen
  \bibfield  {author} {\bibinfo {author} {\bibfnamefont {R.}~\bibnamefont {Khasanov}}, \bibinfo {author} {\bibfnamefont {T.~J.}\ \bibnamefont {Hicken}}, \bibinfo {author} {\bibfnamefont {I.}~\bibnamefont {Plokhikh}}, \bibinfo {author} {\bibfnamefont {V.}~\bibnamefont {Sazgari}}, \bibinfo {author} {\bibfnamefont {L.}~\bibnamefont {Keller}}, \bibinfo {author} {\bibfnamefont {V.}~\bibnamefont {Pomjakushin}}, \bibinfo {author} {\bibfnamefont {M.}~\bibnamefont {Bartkowiak}}, \bibinfo {author} {\bibfnamefont {S.}~\bibnamefont {Kr\'{o}lak}}, \bibinfo {author} {\bibfnamefont {M.~J.}\ \bibnamefont {Winiarski}}, \bibinfo {author} {\bibfnamefont {J.~A.}\ \bibnamefont {Krieger}}, \bibinfo {author} {\bibfnamefont {H.}~\bibnamefont {Luetkens}}, \bibinfo {author} {\bibfnamefont {T.}~\bibnamefont {Klimczuk}}, \bibinfo {author} {\bibfnamefont {D.~J.}\ \bibnamefont {Gawryluk}}, \ and\ \bibinfo {author} {\bibfnamefont {Z.}~\bibnamefont {Guguchia}},\ }\href {https://arxiv.org/abs/2503.04400} {\bibfield  {journal} {\bibinfo
  {journal} {arXiv:2503.04400}\ } (\bibinfo {year} {2025})}\BibitemShut {NoStop}%
\bibitem [{\citenamefont {Xu}\ \emph {et~al.}(2025{\natexlab{b}})\citenamefont {Xu}, \citenamefont {Wang}, \citenamefont {Huo}, \citenamefont {Hu}, \citenamefont {Wu}, \citenamefont {Yue}, \citenamefont {Wu}, \citenamefont {Wang}, \citenamefont {Dong},\ and\ \citenamefont {Wang}}]{Xu2025Collapse}%
  \BibitemOpen
  \bibfield  {author} {\bibinfo {author} {\bibfnamefont {S.}~\bibnamefont {Xu}}, \bibinfo {author} {\bibfnamefont {H.}~\bibnamefont {Wang}}, \bibinfo {author} {\bibfnamefont {M.}~\bibnamefont {Huo}}, \bibinfo {author} {\bibfnamefont {D.}~\bibnamefont {Hu}}, \bibinfo {author} {\bibfnamefont {Q.}~\bibnamefont {Wu}}, \bibinfo {author} {\bibfnamefont {L.}~\bibnamefont {Yue}}, \bibinfo {author} {\bibfnamefont {D.}~\bibnamefont {Wu}}, \bibinfo {author} {\bibfnamefont {M.}~\bibnamefont {Wang}}, \bibinfo {author} {\bibfnamefont {T.}~\bibnamefont {Dong}}, \ and\ \bibinfo {author} {\bibfnamefont {N.}~\bibnamefont {Wang}},\ }\href {https://arxiv.org/abs/2503.05176} {\bibfield  {journal} {\bibinfo  {journal} {arXiv:2503.05176}\ } (\bibinfo {year} {2025}{\natexlab{b}})}\BibitemShut {NoStop}%
\bibitem [{\citenamefont {Jiang}\ \emph {et~al.}(2023)\citenamefont {Jiang}, \citenamefont {Lang}, \citenamefont {Berlijn},\ and\ \citenamefont {Ku}}]{Jiang2022}%
  \BibitemOpen
  \bibfield  {author} {\bibinfo {author} {\bibfnamefont {R.}~\bibnamefont {Jiang}}, \bibinfo {author} {\bibfnamefont {Z.-J.}\ \bibnamefont {Lang}}, \bibinfo {author} {\bibfnamefont {T.}~\bibnamefont {Berlijn}}, \ and\ \bibinfo {author} {\bibfnamefont {W.}~\bibnamefont {Ku}},\ }\href {\doibase 10.1103/PhysRevB.108.155126} {\bibfield  {journal} {\bibinfo  {journal} {Phys. Rev. B}\ }\textbf {\bibinfo {volume} {108}},\ \bibinfo {pages} {155126} (\bibinfo {year} {2023})}\BibitemShut {NoStop}%
\bibitem [{sup()}]{supplementary}%
  \BibitemOpen
  \href@noop {} {}\bibinfo {note} {See Supplementary Material for details of I. Crystal structure of La$_4$Ni$_3$O$_{10}$ under pressure, II. Computational details of the density functional theory calculations, III. Computing band structure in the non-collinear Curie-paramagnetic phase, IV. Atomic Wannier orbitals as a basis for $H^{\text{(Hartree)}}$, V. Extraction of interacting $H^{\text{(Hartree)}}$ from DFT calculations, VI. Leading parameters of $H^{\text{(Hartree)}}$, VII. Obtaining the low-energy effective Hamiltonian via a numerical canonical transformation, VIII. Alternative derivation of effective Hamiltonians, IX. Parameters in deriving $H^{\text{(eV)}}$ from $H^{\text{(Hartree)}}$, X. Rapid growth of $J_{ZZ}$ upon increasing $t^\perp_{pZ}$ through applied pressure, XI. Competition between the Hund's coupling $\tilde{J}_\text{H}$ and superexchange $J_{ZZ}$, XII. Effective sub-eV Hamiltonian with O vacancy; which includes Refs.\,\cite{Singh, Blaha1990, Wei2006, Anisimov1997, Slater,
  chao1977kinetic}.}\BibitemShut {Stop}%
\bibitem [{\citenamefont {Hohenberg}\ and\ \citenamefont {Kohn}(1964)}]{DFT1}%
  \BibitemOpen
  \bibfield  {author} {\bibinfo {author} {\bibfnamefont {P.}~\bibnamefont {Hohenberg}}\ and\ \bibinfo {author} {\bibfnamefont {W.}~\bibnamefont {Kohn}},\ }\href {\doibase 10.1103/PhysRev.136.B864} {\bibfield  {journal} {\bibinfo  {journal} {Phys. Rev.}\ }\textbf {\bibinfo {volume} {136}},\ \bibinfo {pages} {B864} (\bibinfo {year} {1964})}\BibitemShut {NoStop}%
\bibitem [{\citenamefont {Kohn}\ and\ \citenamefont {Sham}(1965)}]{DFT2}%
  \BibitemOpen
  \bibfield  {author} {\bibinfo {author} {\bibfnamefont {W.}~\bibnamefont {Kohn}}\ and\ \bibinfo {author} {\bibfnamefont {L.~J.}\ \bibnamefont {Sham}},\ }\href {\doibase 10.1103/PhysRev.140.A1133} {\bibfield  {journal} {\bibinfo  {journal} {Phys. Rev.}\ }\textbf {\bibinfo {volume} {140}},\ \bibinfo {pages} {A1133} (\bibinfo {year} {1965})}\BibitemShut {NoStop}%
\bibitem [{\citenamefont {Blaha}\ \emph {et~al.}(2001)\citenamefont {Blaha}, \citenamefont {Schwarz}, \citenamefont {Madsen}, \citenamefont {Kvasnicka},\ and\ \citenamefont {Luitz}}]{Blaha2001}%
  \BibitemOpen
  \bibfield  {author} {\bibinfo {author} {\bibfnamefont {P.}~\bibnamefont {Blaha}}, \bibinfo {author} {\bibfnamefont {K.}~\bibnamefont {Schwarz}}, \bibinfo {author} {\bibfnamefont {G.}~\bibnamefont {Madsen}}, \bibinfo {author} {\bibfnamefont {D.}~\bibnamefont {Kvasnicka}}, \ and\ \bibinfo {author} {\bibfnamefont {J.}~\bibnamefont {Luitz}},\ }\href@noop {} {\emph {\bibinfo {title} {WIEN2k: An Augmented Plane Wave plus Local Orbitals Program for Calculating Crystal Properties}}},\ Vol.~\bibinfo {volume} {28}\ (\bibinfo {year} {2001})\BibitemShut {NoStop}%
\bibitem [{\citenamefont {Blaha}\ \emph {et~al.}(2020)\citenamefont {Blaha}, \citenamefont {Schwarz}, \citenamefont {Tran}, \citenamefont {Laskowski}, \citenamefont {Madsen},\ and\ \citenamefont {Marks}}]{Blaha2019}%
  \BibitemOpen
  \bibfield  {author} {\bibinfo {author} {\bibfnamefont {P.}~\bibnamefont {Blaha}}, \bibinfo {author} {\bibfnamefont {K.}~\bibnamefont {Schwarz}}, \bibinfo {author} {\bibfnamefont {F.}~\bibnamefont {Tran}}, \bibinfo {author} {\bibfnamefont {R.}~\bibnamefont {Laskowski}}, \bibinfo {author} {\bibfnamefont {G.~K.~H.}\ \bibnamefont {Madsen}}, \ and\ \bibinfo {author} {\bibfnamefont {L.~D.}\ \bibnamefont {Marks}},\ }\href {\doibase 10.1063/1.5143061} {\bibfield  {journal} {\bibinfo  {journal} {The Journal of Chemical Physics}\ }\textbf {\bibinfo {volume} {152}},\ \bibinfo {pages} {074101} (\bibinfo {year} {2020})}\BibitemShut {NoStop}%
\bibitem [{\citenamefont {Anisimov}\ \emph {et~al.}(1993)\citenamefont {Anisimov}, \citenamefont {Solovyev}, \citenamefont {Korotin}, \citenamefont {Czy\ifmmode~\dot{z}\else \.{z}\fi{}yk},\ and\ \citenamefont {Sawatzky}}]{Anisimov1993}%
  \BibitemOpen
  \bibfield  {author} {\bibinfo {author} {\bibfnamefont {V.~I.}\ \bibnamefont {Anisimov}}, \bibinfo {author} {\bibfnamefont {I.~V.}\ \bibnamefont {Solovyev}}, \bibinfo {author} {\bibfnamefont {M.~A.}\ \bibnamefont {Korotin}}, \bibinfo {author} {\bibfnamefont {M.~T.}\ \bibnamefont {Czy\ifmmode~\dot{z}\else \.{z}\fi{}yk}}, \ and\ \bibinfo {author} {\bibfnamefont {G.~A.}\ \bibnamefont {Sawatzky}},\ }\href {\doibase 10.1103/PhysRevB.48.16929} {\bibfield  {journal} {\bibinfo  {journal} {Phys. Rev. B}\ }\textbf {\bibinfo {volume} {48}},\ \bibinfo {pages} {16929} (\bibinfo {year} {1993})}\BibitemShut {NoStop}%
\bibitem [{\citenamefont {Liechtenstein}\ \emph {et~al.}(1995)\citenamefont {Liechtenstein}, \citenamefont {Anisimov},\ and\ \citenamefont {Zaanen}}]{Liechtenstein1995}%
  \BibitemOpen
  \bibfield  {author} {\bibinfo {author} {\bibfnamefont {A.~I.}\ \bibnamefont {Liechtenstein}}, \bibinfo {author} {\bibfnamefont {V.~I.}\ \bibnamefont {Anisimov}}, \ and\ \bibinfo {author} {\bibfnamefont {J.}~\bibnamefont {Zaanen}},\ }\href {\doibase 10.1103/PhysRevB.52.R5467} {\bibfield  {journal} {\bibinfo  {journal} {Phys. Rev. B}\ }\textbf {\bibinfo {volume} {52}},\ \bibinfo {pages} {R5467} (\bibinfo {year} {1995})}\BibitemShut {NoStop}%
\bibitem [{\citenamefont {Dudarev}\ \emph {et~al.}(1998)\citenamefont {Dudarev}, \citenamefont {Botton}, \citenamefont {Savrasov}, \citenamefont {Humphreys},\ and\ \citenamefont {Sutton}}]{Dudarev1998}%
  \BibitemOpen
  \bibfield  {author} {\bibinfo {author} {\bibfnamefont {S.~L.}\ \bibnamefont {Dudarev}}, \bibinfo {author} {\bibfnamefont {G.~A.}\ \bibnamefont {Botton}}, \bibinfo {author} {\bibfnamefont {S.~Y.}\ \bibnamefont {Savrasov}}, \bibinfo {author} {\bibfnamefont {C.~J.}\ \bibnamefont {Humphreys}}, \ and\ \bibinfo {author} {\bibfnamefont {A.~P.}\ \bibnamefont {Sutton}},\ }\href {\doibase 10.1103/PhysRevB.57.1505} {\bibfield  {journal} {\bibinfo  {journal} {Phys. Rev. B}\ }\textbf {\bibinfo {volume} {57}},\ \bibinfo {pages} {1505} (\bibinfo {year} {1998})}\BibitemShut {NoStop}%
\bibitem [{\citenamefont {Jiang}\ \emph {et~al.}(2024{\natexlab{b}})\citenamefont {Jiang}, \citenamefont {Gu},\ and\ \citenamefont {Ku}}]{Jiang2024Interactionan}%
  \BibitemOpen
  \bibfield  {author} {\bibinfo {author} {\bibfnamefont {R.}~\bibnamefont {Jiang}}, \bibinfo {author} {\bibfnamefont {F.}~\bibnamefont {Gu}}, \ and\ \bibinfo {author} {\bibfnamefont {W.}~\bibnamefont {Ku}},\ }\href {https://arxiv.org/abs/2407.03319} {\bibfield  {journal} {\bibinfo  {journal} {arXiv:2407.03319}\ } (\bibinfo {year} {2024}{\natexlab{b}})}\BibitemShut {NoStop}%
\bibitem [{\citenamefont {Ku}\ \emph {et~al.}(2002)\citenamefont {Ku}, \citenamefont {Rosner}, \citenamefont {Pickett},\ and\ \citenamefont {Scalettar}}]{Wei2002}%
  \BibitemOpen
  \bibfield  {author} {\bibinfo {author} {\bibfnamefont {W.}~\bibnamefont {Ku}}, \bibinfo {author} {\bibfnamefont {H.}~\bibnamefont {Rosner}}, \bibinfo {author} {\bibfnamefont {W.~E.}\ \bibnamefont {Pickett}}, \ and\ \bibinfo {author} {\bibfnamefont {R.~T.}\ \bibnamefont {Scalettar}},\ }\href {\doibase 10.1103/PhysRevLett.89.167204} {\bibfield  {journal} {\bibinfo  {journal} {Phys. Rev. Lett.}\ }\textbf {\bibinfo {volume} {89}},\ \bibinfo {pages} {167204} (\bibinfo {year} {2002})}\BibitemShut {NoStop}%
\bibitem [{\citenamefont {Marzari}\ and\ \citenamefont {Vanderbilt}(1997)}]{Marzari1997}%
  \BibitemOpen
  \bibfield  {author} {\bibinfo {author} {\bibfnamefont {N.}~\bibnamefont {Marzari}}\ and\ \bibinfo {author} {\bibfnamefont {D.}~\bibnamefont {Vanderbilt}},\ }\href {\doibase 10.1103/PhysRevB.56.12847} {\bibfield  {journal} {\bibinfo  {journal} {Phys. Rev. B}\ }\textbf {\bibinfo {volume} {56}},\ \bibinfo {pages} {12847} (\bibinfo {year} {1997})}\BibitemShut {NoStop}%
\bibitem [{\citenamefont {Ku}\ \emph {et~al.}(2010)\citenamefont {Ku}, \citenamefont {Berlijn},\ and\ \citenamefont {Lee}}]{Wei2010}%
  \BibitemOpen
  \bibfield  {author} {\bibinfo {author} {\bibfnamefont {W.}~\bibnamefont {Ku}}, \bibinfo {author} {\bibfnamefont {T.}~\bibnamefont {Berlijn}}, \ and\ \bibinfo {author} {\bibfnamefont {C.-C.}\ \bibnamefont {Lee}},\ }\href {\doibase 10.1103/PhysRevLett.104.216401} {\bibfield  {journal} {\bibinfo  {journal} {Phys. Rev. Lett.}\ }\textbf {\bibinfo {volume} {104}},\ \bibinfo {pages} {216401} (\bibinfo {year} {2010})}\BibitemShut {NoStop}%
\bibitem [{\citenamefont {Zhang}\ and\ \citenamefont {Rice}(1988)}]{Zhang1988}%
  \BibitemOpen
  \bibfield  {author} {\bibinfo {author} {\bibfnamefont {F.~C.}\ \bibnamefont {Zhang}}\ and\ \bibinfo {author} {\bibfnamefont {T.~M.}\ \bibnamefont {Rice}},\ }\href {\doibase 10.1103/PhysRevB.37.3759} {\bibfield  {journal} {\bibinfo  {journal} {Phys. Rev. B}\ }\textbf {\bibinfo {volume} {37}},\ \bibinfo {pages} {3759} (\bibinfo {year} {1988})}\BibitemShut {NoStop}%
\bibitem [{not()}]{note_hole_impact}%
  \BibitemOpen
  \href@noop {} {}\bibinfo {note} {Introduction of the remaining ligand hole will not qualitatively affect our study, since for the local Ni-O-Ni-O-Ni component, its main effect is to fluctuate the $d_{x^2-y^2}$ orbital, effectively reducing the Hund's coupling and thus enhancing the main physics found at high-pressure.}\BibitemShut {Stop}%
\bibitem [{\citenamefont {Ogata}\ and\ \citenamefont {Fukuyama}(2008)}]{Ogata_2008}%
  \BibitemOpen
  \bibfield  {author} {\bibinfo {author} {\bibfnamefont {M.}~\bibnamefont {Ogata}}\ and\ \bibinfo {author} {\bibfnamefont {H.}~\bibnamefont {Fukuyama}},\ }\href {\doibase 10.1088/0034-4885/71/3/036501} {\bibfield  {journal} {\bibinfo  {journal} {Reports on Progress in Physics}\ }\textbf {\bibinfo {volume} {71}},\ \bibinfo {pages} {036501} (\bibinfo {year} {2008})}\BibitemShut {NoStop}%
\bibitem [{\citenamefont {White}(2002)}]{White2002}%
  \BibitemOpen
  \bibfield  {author} {\bibinfo {author} {\bibfnamefont {S.}~\bibnamefont {White}},\ }\href {https://doi.org/10.1063/1.1508370} {\bibfield  {journal} {\bibinfo  {journal} {The Journal of Chemical Physics}\ }\textbf {\bibinfo {volume} {117}},\ \bibinfo {pages} {7472} (\bibinfo {year} {2002})}\BibitemShut {NoStop}%
\bibitem [{\citenamefont {Schrieffer}\ and\ \citenamefont {Wolff}(1966)}]{Schrieffer1966}%
  \BibitemOpen
  \bibfield  {author} {\bibinfo {author} {\bibfnamefont {J.~R.}\ \bibnamefont {Schrieffer}}\ and\ \bibinfo {author} {\bibfnamefont {P.~A.}\ \bibnamefont {Wolff}},\ }\href {\doibase 10.1103/PhysRev.149.491} {\bibfield  {journal} {\bibinfo  {journal} {Phys. Rev.}\ }\textbf {\bibinfo {volume} {149}},\ \bibinfo {pages} {491} (\bibinfo {year} {1966})}\BibitemShut {NoStop}%
\bibitem [{\citenamefont {Zaanen}\ and\ \citenamefont {Ole\ifmmode~\acute{s}\else \'{s}\fi{}}(1988)}]{Zaanen1988}%
  \BibitemOpen
  \bibfield  {author} {\bibinfo {author} {\bibfnamefont {J.}~\bibnamefont {Zaanen}}\ and\ \bibinfo {author} {\bibfnamefont {A.~M.}\ \bibnamefont {Ole\ifmmode~\acute{s}\else \'{s}\fi{}}},\ }\href {\doibase 10.1103/PhysRevB.37.9423} {\bibfield  {journal} {\bibinfo  {journal} {Phys. Rev. B}\ }\textbf {\bibinfo {volume} {37}},\ \bibinfo {pages} {9423} (\bibinfo {year} {1988})}\BibitemShut {NoStop}%
\bibitem [{\citenamefont {Yin}\ and\ \citenamefont {Ku}(2009)}]{Yin2009}%
  \BibitemOpen
  \bibfield  {author} {\bibinfo {author} {\bibfnamefont {W.-G.}\ \bibnamefont {Yin}}\ and\ \bibinfo {author} {\bibfnamefont {W.}~\bibnamefont {Ku}},\ }\href {\doibase 10.1103/PhysRevB.79.214512} {\bibfield  {journal} {\bibinfo  {journal} {Phys. Rev. B}\ }\textbf {\bibinfo {volume} {79}},\ \bibinfo {pages} {214512} (\bibinfo {year} {2009})}\BibitemShut {NoStop}%
\bibitem [{\citenamefont {Anderson}(1950)}]{Anderson1950}%
  \BibitemOpen
  \bibfield  {author} {\bibinfo {author} {\bibfnamefont {P.~W.}\ \bibnamefont {Anderson}},\ }\href {\doibase 10.1103/PhysRev.79.350} {\bibfield  {journal} {\bibinfo  {journal} {Phys. Rev.}\ }\textbf {\bibinfo {volume} {79}},\ \bibinfo {pages} {350} (\bibinfo {year} {1950})}\BibitemShut {NoStop}%
\bibitem [{\citenamefont {Tan}\ \emph {et~al.}(2022)\citenamefont {Tan}, \citenamefont {Zhang}, \citenamefont {Zou}, \citenamefont {dos Santos}, \citenamefont {Hu}, \citenamefont {Yao}, \citenamefont {Mao}, \citenamefont {Ke},\ and\ \citenamefont {Ku}}]{Tan2022}%
  \BibitemOpen
  \bibfield  {author} {\bibinfo {author} {\bibfnamefont {Y.}~\bibnamefont {Tan}}, \bibinfo {author} {\bibfnamefont {T.}~\bibnamefont {Zhang}}, \bibinfo {author} {\bibfnamefont {T.}~\bibnamefont {Zou}}, \bibinfo {author} {\bibfnamefont {A.~M.}\ \bibnamefont {dos Santos}}, \bibinfo {author} {\bibfnamefont {J.}~\bibnamefont {Hu}}, \bibinfo {author} {\bibfnamefont {D.-X.}\ \bibnamefont {Yao}}, \bibinfo {author} {\bibfnamefont {Z.~Q.}\ \bibnamefont {Mao}}, \bibinfo {author} {\bibfnamefont {X.}~\bibnamefont {Ke}}, \ and\ \bibinfo {author} {\bibfnamefont {W.}~\bibnamefont {Ku}},\ }\href {\doibase 10.1103/PhysRevResearch.4.033115} {\bibfield  {journal} {\bibinfo  {journal} {Phys. Rev. Res.}\ }\textbf {\bibinfo {volume} {4}},\ \bibinfo {pages} {033115} (\bibinfo {year} {2022})}\BibitemShut {NoStop}%
\bibitem [{\citenamefont {Wu}\ \emph {et~al.}(2020)\citenamefont {Wu}, \citenamefont {Di~Sante}, \citenamefont {Schwemmer}, \citenamefont {Hanke}, \citenamefont {Hwang}, \citenamefont {Raghu},\ and\ \citenamefont {Thomale}}]{Wu2019Robust}%
  \BibitemOpen
  \bibfield  {author} {\bibinfo {author} {\bibfnamefont {X.}~\bibnamefont {Wu}}, \bibinfo {author} {\bibfnamefont {D.}~\bibnamefont {Di~Sante}}, \bibinfo {author} {\bibfnamefont {T.}~\bibnamefont {Schwemmer}}, \bibinfo {author} {\bibfnamefont {W.}~\bibnamefont {Hanke}}, \bibinfo {author} {\bibfnamefont {H.~Y.}\ \bibnamefont {Hwang}}, \bibinfo {author} {\bibfnamefont {S.}~\bibnamefont {Raghu}}, \ and\ \bibinfo {author} {\bibfnamefont {R.}~\bibnamefont {Thomale}},\ }\href {\doibase 10.1103/PhysRevB.101.060504} {\bibfield  {journal} {\bibinfo  {journal} {Phys. Rev. B}\ }\textbf {\bibinfo {volume} {101}},\ \bibinfo {pages} {060504} (\bibinfo {year} {2020})}\BibitemShut {NoStop}%
\bibitem [{\citenamefont {Lechermann}\ \emph {et~al.}(2023)\citenamefont {Lechermann}, \citenamefont {Gondolf}, \citenamefont {B\"otzel},\ and\ \citenamefont {Eremin}}]{Lechermann2023Electronic}%
  \BibitemOpen
  \bibfield  {author} {\bibinfo {author} {\bibfnamefont {F.}~\bibnamefont {Lechermann}}, \bibinfo {author} {\bibfnamefont {J.}~\bibnamefont {Gondolf}}, \bibinfo {author} {\bibfnamefont {S.}~\bibnamefont {B\"otzel}}, \ and\ \bibinfo {author} {\bibfnamefont {I.~M.}\ \bibnamefont {Eremin}},\ }\href {\doibase 10.1103/PhysRevB.108.L201121} {\bibfield  {journal} {\bibinfo  {journal} {Phys. Rev. B}\ }\textbf {\bibinfo {volume} {108}},\ \bibinfo {pages} {L201121} (\bibinfo {year} {2023})}\BibitemShut {NoStop}%
\bibitem [{\citenamefont {Jiang}\ \emph {et~al.}(2024{\natexlab{c}})\citenamefont {Jiang}, \citenamefont {Wang},\ and\ \citenamefont {Zhang}}]{Jiang2024High-Temperature}%
  \BibitemOpen
  \bibfield  {author} {\bibinfo {author} {\bibfnamefont {K.}~\bibnamefont {Jiang}}, \bibinfo {author} {\bibfnamefont {Z.}~\bibnamefont {Wang}}, \ and\ \bibinfo {author} {\bibfnamefont {F.-C.}\ \bibnamefont {Zhang}},\ }\href {\doibase 10.1088/0256-307x/41/1/017402} {\bibfield  {journal} {\bibinfo  {journal} {Chinese Physics Letters}\ }\textbf {\bibinfo {volume} {41}},\ \bibinfo {pages} {017402} (\bibinfo {year} {2024}{\natexlab{c}})}\BibitemShut {NoStop}%
\bibitem [{\citenamefont {Kitaev}(2006)}]{kitaev2006}%
  \BibitemOpen
  \bibfield  {author} {\bibinfo {author} {\bibfnamefont {A.}~\bibnamefont {Kitaev}},\ }\href {\doibase https://doi.org/10.1016/j.aop.2005.10.005} {\bibfield  {journal} {\bibinfo  {journal} {Annals of Physics}\ }\textbf {\bibinfo {volume} {321}},\ \bibinfo {pages} {2} (\bibinfo {year} {2006})},\ \bibinfo {note} {january Special Issue}\BibitemShut {NoStop}%
\bibitem [{\citenamefont {Wang}\ \emph {et~al.}(2024{\natexlab{b}})\citenamefont {Wang}, \citenamefont {Wang}, \citenamefont {Shen}, \citenamefont {Hou}, \citenamefont {Luo}, \citenamefont {Ma}, \citenamefont {Yang}, \citenamefont {Shi}, \citenamefont {Dou}, \citenamefont {Feng}, \citenamefont {Yang}, \citenamefont {Shi}, \citenamefont {Ren}, \citenamefont {Ma}, \citenamefont {Yang}, \citenamefont {Liu}, \citenamefont {Liu}, \citenamefont {Zhang}, \citenamefont {Dong},\ and\ \citenamefont {Cheng}}]{Wang2024Bulk}%
  \BibitemOpen
  \bibfield  {author} {\bibinfo {author} {\bibfnamefont {N.}~\bibnamefont {Wang}}, \bibinfo {author} {\bibfnamefont {G.}~\bibnamefont {Wang}}, \bibinfo {author} {\bibfnamefont {X.}~\bibnamefont {Shen}}, \bibinfo {author} {\bibfnamefont {J.}~\bibnamefont {Hou}}, \bibinfo {author} {\bibfnamefont {J.}~\bibnamefont {Luo}}, \bibinfo {author} {\bibfnamefont {X.}~\bibnamefont {Ma}}, \bibinfo {author} {\bibfnamefont {H.}~\bibnamefont {Yang}}, \bibinfo {author} {\bibfnamefont {L.}~\bibnamefont {Shi}}, \bibinfo {author} {\bibfnamefont {J.}~\bibnamefont {Dou}}, \bibinfo {author} {\bibfnamefont {J.}~\bibnamefont {Feng}}, \bibinfo {author} {\bibfnamefont {J.}~\bibnamefont {Yang}}, \bibinfo {author} {\bibfnamefont {Y.}~\bibnamefont {Shi}}, \bibinfo {author} {\bibfnamefont {Z.}~\bibnamefont {Ren}}, \bibinfo {author} {\bibfnamefont {H.}~\bibnamefont {Ma}}, \bibinfo {author} {\bibfnamefont {P.}~\bibnamefont {Yang}}, \bibinfo {author} {\bibfnamefont {Z.}~\bibnamefont {Liu}}, \bibinfo {author} {\bibfnamefont {Y.}~\bibnamefont
  {Liu}}, \bibinfo {author} {\bibfnamefont {H.}~\bibnamefont {Zhang}}, \bibinfo {author} {\bibfnamefont {X.}~\bibnamefont {Dong}}, \ and\ \bibinfo {author} {\bibfnamefont {J.-G.}\ \bibnamefont {Cheng}},\ }\href {\doibase 10.1038/s41586-024-07996-8} {\bibfield  {journal} {\bibinfo  {journal} {Nature}\ }\textbf {\bibinfo {volume} {634}},\ \bibinfo {pages} {579} (\bibinfo {year} {2024}{\natexlab{b}})}\BibitemShut {NoStop}%
\bibitem [{\citenamefont {Li}\ \emph {et~al.}(2025{\natexlab{c}})\citenamefont {Li}, \citenamefont {Xing}, \citenamefont {Peng}, \citenamefont {Dou}, \citenamefont {Guo}, \citenamefont {Ma}, \citenamefont {Zhang}, \citenamefont {Wang}, \citenamefont {Luo}, \citenamefont {Yang}, \citenamefont {Zhang}, \citenamefont {Chang}, \citenamefont {Chen}, \citenamefont {Cai}, \citenamefont {Cheng}, \citenamefont {Wang}, \citenamefont {Zeng}, \citenamefont {Zheng}, \citenamefont {Zhou}, \citenamefont {Zeng}, \citenamefont {Tao},\ and\ \citenamefont {Zhang}}]{Li2025Ambient}%
  \BibitemOpen
  \bibfield  {author} {\bibinfo {author} {\bibfnamefont {F.}~\bibnamefont {Li}}, \bibinfo {author} {\bibfnamefont {Z.}~\bibnamefont {Xing}}, \bibinfo {author} {\bibfnamefont {D.}~\bibnamefont {Peng}}, \bibinfo {author} {\bibfnamefont {J.}~\bibnamefont {Dou}}, \bibinfo {author} {\bibfnamefont {N.}~\bibnamefont {Guo}}, \bibinfo {author} {\bibfnamefont {L.}~\bibnamefont {Ma}}, \bibinfo {author} {\bibfnamefont {Y.}~\bibnamefont {Zhang}}, \bibinfo {author} {\bibfnamefont {L.}~\bibnamefont {Wang}}, \bibinfo {author} {\bibfnamefont {J.}~\bibnamefont {Luo}}, \bibinfo {author} {\bibfnamefont {J.}~\bibnamefont {Yang}}, \bibinfo {author} {\bibfnamefont {J.}~\bibnamefont {Zhang}}, \bibinfo {author} {\bibfnamefont {T.}~\bibnamefont {Chang}}, \bibinfo {author} {\bibfnamefont {Y.-S.}\ \bibnamefont {Chen}}, \bibinfo {author} {\bibfnamefont {W.}~\bibnamefont {Cai}}, \bibinfo {author} {\bibfnamefont {J.}~\bibnamefont {Cheng}}, \bibinfo {author} {\bibfnamefont {Y.}~\bibnamefont {Wang}}, \bibinfo {author} {\bibfnamefont
  {Z.}~\bibnamefont {Zeng}}, \bibinfo {author} {\bibfnamefont {Q.}~\bibnamefont {Zheng}}, \bibinfo {author} {\bibfnamefont {R.}~\bibnamefont {Zhou}}, \bibinfo {author} {\bibfnamefont {Q.}~\bibnamefont {Zeng}}, \bibinfo {author} {\bibfnamefont {X.}~\bibnamefont {Tao}}, \ and\ \bibinfo {author} {\bibfnamefont {J.}~\bibnamefont {Zhang}},\ }\href {https://arxiv.org/abs/2501.14584} {\bibfield  {journal} {\bibinfo  {journal} {arXiv:2501.14584}\ } (\bibinfo {year} {2025}{\natexlab{c}})}\BibitemShut {NoStop}%
\bibitem [{\citenamefont {Shi}\ \emph {et~al.}(2025)\citenamefont {Shi}, \citenamefont {Peng}, \citenamefont {Fan}, \citenamefont {Xing}, \citenamefont {Yang}, \citenamefont {Wang}, \citenamefont {Li}, \citenamefont {Wu}, \citenamefont {Du}, \citenamefont {Ge}, \citenamefont {Zeng}, \citenamefont {Zeng}, \citenamefont {Ying}, \citenamefont {Wu},\ and\ \citenamefont {Chen}}]{Mengzhu2025Superconductivity}%
  \BibitemOpen
  \bibfield  {author} {\bibinfo {author} {\bibfnamefont {M.}~\bibnamefont {Shi}}, \bibinfo {author} {\bibfnamefont {D.}~\bibnamefont {Peng}}, \bibinfo {author} {\bibfnamefont {K.}~\bibnamefont {Fan}}, \bibinfo {author} {\bibfnamefont {Z.}~\bibnamefont {Xing}}, \bibinfo {author} {\bibfnamefont {S.}~\bibnamefont {Yang}}, \bibinfo {author} {\bibfnamefont {Y.}~\bibnamefont {Wang}}, \bibinfo {author} {\bibfnamefont {H.}~\bibnamefont {Li}}, \bibinfo {author} {\bibfnamefont {R.}~\bibnamefont {Wu}}, \bibinfo {author} {\bibfnamefont {M.}~\bibnamefont {Du}}, \bibinfo {author} {\bibfnamefont {B.}~\bibnamefont {Ge}}, \bibinfo {author} {\bibfnamefont {Z.}~\bibnamefont {Zeng}}, \bibinfo {author} {\bibfnamefont {Q.}~\bibnamefont {Zeng}}, \bibinfo {author} {\bibfnamefont {J.}~\bibnamefont {Ying}}, \bibinfo {author} {\bibfnamefont {T.}~\bibnamefont {Wu}}, \ and\ \bibinfo {author} {\bibfnamefont {X.}~\bibnamefont {Chen}},\ }\href {https://arxiv.org/abs/2502.01018} {\bibfield  {journal} {\bibinfo  {journal} {arXiv:2502.01018}\
  } (\bibinfo {year} {2025})}\BibitemShut {NoStop}%
\bibitem [{\citenamefont {Yatoo}\ \emph {et~al.}(2023)\citenamefont {Yatoo}, \citenamefont {Seymour},\ and\ \citenamefont {Skinner}}]{Yatoo2023Neutron}%
  \BibitemOpen
  \bibfield  {author} {\bibinfo {author} {\bibfnamefont {M.~A.}\ \bibnamefont {Yatoo}}, \bibinfo {author} {\bibfnamefont {I.~D.}\ \bibnamefont {Seymour}}, \ and\ \bibinfo {author} {\bibfnamefont {S.~J.}\ \bibnamefont {Skinner}},\ }\href {\doibase 10.1039/D3RA01772A} {\bibfield  {journal} {\bibinfo  {journal} {RSC Adv.}\ }\textbf {\bibinfo {volume} {13}},\ \bibinfo {pages} {13786} (\bibinfo {year} {2023})}\BibitemShut {NoStop}%
\bibitem [{\citenamefont {Puphal}\ \emph {et~al.}(2024)\citenamefont {Puphal}, \citenamefont {Reiss}, \citenamefont {Enderlein}, \citenamefont {Wu}, \citenamefont {Khaliullin}, \citenamefont {Sundaramurthy}, \citenamefont {Priessnitz}, \citenamefont {Knauft}, \citenamefont {Suthar}, \citenamefont {Richter}, \citenamefont {Isobe}, \citenamefont {van Aken}, \citenamefont {Takagi}, \citenamefont {Keimer}, \citenamefont {Suyolcu}, \citenamefont {Wehinger}, \citenamefont {Hansmann},\ and\ \citenamefont {Hepting}}]{Puphal2024Unconventional}%
  \BibitemOpen
  \bibfield  {author} {\bibinfo {author} {\bibfnamefont {P.}~\bibnamefont {Puphal}}, \bibinfo {author} {\bibfnamefont {P.}~\bibnamefont {Reiss}}, \bibinfo {author} {\bibfnamefont {N.}~\bibnamefont {Enderlein}}, \bibinfo {author} {\bibfnamefont {Y.-M.}\ \bibnamefont {Wu}}, \bibinfo {author} {\bibfnamefont {G.}~\bibnamefont {Khaliullin}}, \bibinfo {author} {\bibfnamefont {V.}~\bibnamefont {Sundaramurthy}}, \bibinfo {author} {\bibfnamefont {T.}~\bibnamefont {Priessnitz}}, \bibinfo {author} {\bibfnamefont {M.}~\bibnamefont {Knauft}}, \bibinfo {author} {\bibfnamefont {A.}~\bibnamefont {Suthar}}, \bibinfo {author} {\bibfnamefont {L.}~\bibnamefont {Richter}}, \bibinfo {author} {\bibfnamefont {M.}~\bibnamefont {Isobe}}, \bibinfo {author} {\bibfnamefont {P.~A.}\ \bibnamefont {van Aken}}, \bibinfo {author} {\bibfnamefont {H.}~\bibnamefont {Takagi}}, \bibinfo {author} {\bibfnamefont {B.}~\bibnamefont {Keimer}}, \bibinfo {author} {\bibfnamefont {Y.~E.}\ \bibnamefont {Suyolcu}}, \bibinfo {author} {\bibfnamefont
  {B.}~\bibnamefont {Wehinger}}, \bibinfo {author} {\bibfnamefont {P.}~\bibnamefont {Hansmann}}, \ and\ \bibinfo {author} {\bibfnamefont {M.}~\bibnamefont {Hepting}},\ }\href {\doibase 10.1103/PhysRevLett.133.146002} {\bibfield  {journal} {\bibinfo  {journal} {Phys. Rev. Lett.}\ }\textbf {\bibinfo {volume} {133}},\ \bibinfo {pages} {146002} (\bibinfo {year} {2024})}\BibitemShut {NoStop}%
\bibitem [{\citenamefont {Singh}(2006)}]{Singh}%
  \BibitemOpen
  \bibfield  {author} {\bibinfo {author} {\bibfnamefont {D.~J.}\ \bibnamefont {Singh}},\ }\href@noop {} {\emph {\bibinfo {title} {Planewaves, Pseudopotentials and the LAPW Method}}}\ (\bibinfo  {publisher} {Springer New York, NY},\ \bibinfo {address} {New York},\ \bibinfo {year} {2006})\BibitemShut {NoStop}%
\bibitem [{\citenamefont {Blaha}\ \emph {et~al.}(1990)\citenamefont {Blaha}, \citenamefont {Schwarz}, \citenamefont {Sorantin},\ and\ \citenamefont {Trickey}}]{Blaha1990}%
  \BibitemOpen
  \bibfield  {author} {\bibinfo {author} {\bibfnamefont {P.}~\bibnamefont {Blaha}}, \bibinfo {author} {\bibfnamefont {K.}~\bibnamefont {Schwarz}}, \bibinfo {author} {\bibfnamefont {P.}~\bibnamefont {Sorantin}}, \ and\ \bibinfo {author} {\bibfnamefont {S.}~\bibnamefont {Trickey}},\ }\href {\doibase https://doi.org/10.1016/0010-4655(90)90187-6} {\bibfield  {journal} {\bibinfo  {journal} {Computer Physics Communications}\ }\textbf {\bibinfo {volume} {59}},\ \bibinfo {pages} {399} (\bibinfo {year} {1990})}\BibitemShut {NoStop}%
\bibitem [{\citenamefont {Yin}\ \emph {et~al.}(2006)\citenamefont {Yin}, \citenamefont {Volja},\ and\ \citenamefont {Ku}}]{Wei2006}%
  \BibitemOpen
  \bibfield  {author} {\bibinfo {author} {\bibfnamefont {W.-G.}\ \bibnamefont {Yin}}, \bibinfo {author} {\bibfnamefont {D.}~\bibnamefont {Volja}}, \ and\ \bibinfo {author} {\bibfnamefont {W.}~\bibnamefont {Ku}},\ }\href {\doibase 10.1103/PhysRevLett.96.116405} {\bibfield  {journal} {\bibinfo  {journal} {Phys. Rev. Lett.}\ }\textbf {\bibinfo {volume} {96}},\ \bibinfo {pages} {116405} (\bibinfo {year} {2006})}\BibitemShut {NoStop}%
\bibitem [{\citenamefont {Anisimov}\ \emph {et~al.}(1997)\citenamefont {Anisimov}, \citenamefont {Aryasetiawan},\ and\ \citenamefont {Lichtenstein}}]{Anisimov1997}%
  \BibitemOpen
  \bibfield  {author} {\bibinfo {author} {\bibfnamefont {V.~I.}\ \bibnamefont {Anisimov}}, \bibinfo {author} {\bibfnamefont {F.}~\bibnamefont {Aryasetiawan}}, \ and\ \bibinfo {author} {\bibfnamefont {A.~I.}\ \bibnamefont {Lichtenstein}},\ }\href {\doibase 10.1088/0953-8984/9/4/002} {\bibfield  {journal} {\bibinfo  {journal} {Journal of Physics: Condensed Matter}\ }\textbf {\bibinfo {volume} {9}},\ \bibinfo {pages} {767} (\bibinfo {year} {1997})}\BibitemShut {NoStop}%
\bibitem [{\citenamefont {Slater}(1974)}]{Slater}%
  \BibitemOpen
  \bibfield  {author} {\bibinfo {author} {\bibfnamefont {J.}~\bibnamefont {Slater}},\ }\href@noop {} {\emph {\bibinfo {title} {Quantum Theory of Molecules and Solids}}}\ (\bibinfo  {publisher} {Mcgram-Hill},\ \bibinfo {address} {New York},\ \bibinfo {year} {1974})\BibitemShut {NoStop}%
\bibitem [{\citenamefont {Chao}\ \emph {et~al.}(1977)\citenamefont {Chao}, \citenamefont {Spalek},\ and\ \citenamefont {Oles}}]{chao1977kinetic}%
  \BibitemOpen
  \bibfield  {author} {\bibinfo {author} {\bibfnamefont {K.~A.}\ \bibnamefont {Chao}}, \bibinfo {author} {\bibfnamefont {J.}~\bibnamefont {Spalek}}, \ and\ \bibinfo {author} {\bibfnamefont {A.~M.}\ \bibnamefont {Oles}},\ }\href {\doibase 10.1088/0022-3719/10/10/002} {\bibfield  {journal} {\bibinfo  {journal} {Journal of Physics C: Solid State Physics}\ }\textbf {\bibinfo {volume} {10}},\ \bibinfo {pages} {L271} (\bibinfo {year} {1977})}\BibitemShut {NoStop}%
\end{thebibliography}%
\end{document}